\documentclass[journal]{IEEEtran}
\ifCLASSINFOpdf
\else
\fi 
\usepackage{ragged2e}
\usepackage{floatrow}
\usepackage{amsthm}
\usepackage[utf8]{inputenc}
\usepackage[english]{babel}
\usepackage{multicol}
\usepackage{epsfig}
\usepackage{epstopdf}
\usepackage{float}
\usepackage{perpage} 
\usepackage{array}
\usepackage{graphicx}
\usepackage{amsfonts}
\usepackage{amssymb}
\usepackage{comment}
\usepackage{hyperref} 
\usepackage{subcaption} 
\usepackage{algorithm}
\usepackage{algpseudocode} 
\usepackage{cite}
\usepackage{array} 
\usepackage[utf8]{inputenc}
\usepackage{amsmath}
\usepackage{amssymb}
\usepackage{caption}
\newtheorem{Remark}{Remark}
\newtheorem{theorem}{Theorem}
\newtheorem{Proposition}{Proposition}
\newtheorem{lemma}{Lemma} 
\usepackage{tabularx}
\usepackage{amssymb} 
\usepackage{pifont} 
\begin{document}

\title{{Analysis and Optimization of RIS-Assisted Cell-Free Massive MIMO NOMA Systems}}

\author{Malay Chakraborty, \textit{Student Member, IEEE}, Ekant Sharma, \textit{Senior Member, IEEE}, Himal A. Suraweera, \textit{Senior Member, IEEE}, and Hien Quoc Ngo, \textit{Senior Member, IEEE} \\[-30pt]
\thanks{Ekant Sharma would like to acknowledge the financial support under the project “ITB-1983-ECD” from IIITB COMET Foundation and SRG/2022/000154 from the Science and Engineering Research Board (SERB), Government of India. Hien Quoc Ngo would like to acknowledge the financial support under the U.K. Research and Innovation Future Leaders Fellowships under Grant MR/X010635/1, and a research grant from the Department for the Economy Northern Ireland under the US-Ireland R\&D Partnership Programme. This paper was presented in part at the IEEE Globecom Workshops (GC Wkshps), Rio de Janeiro, Brazil, in 2022.}
\thanks{Malay Chakraborty and Ekant Sharma are with the Department of Electronics and Communication Engineering,  Indian Institute of Technology Roorkee, Roorkee 247667, India (e-mail: malay\_c@ece.iitr.ac.in; ekant@ece.iitr.ac.in).} 
\thanks{Himal A. Suraweera is with the Department of Electrical and Electronic Engineering, University of Peradeniya, Peradeniya 20400, Sri Lanka (e-mail: himal@ee.pdn.ac.lk).} 
\thanks{Hien Quoc Ngo is with the Centre for Wireless Innovation (CWI), Queen’s University Belfast, BT3 9DT Belfast, U.K. (e-mail: hien.ngo@qub.ac.uk).}
\vspace{-0pt}
}

\maketitle
\begin{abstract}
We consider a reconfigurable intelligent surface (RIS) assisted cell-free massive multiple-input multiple-output non-orthogonal multiple access (NOMA) system, where each access point (AP) serves all the users with the aid of the RIS. We practically model the system by considering imperfect instantaneous channel state information (CSI) and employing imperfect successive interference cancellation at the users' end. We first obtain the channel estimates using linear minimum mean square error approach considering the spatial correlation at the RIS and then derive a closed-form downlink spectral efficiency (SE) expression using the statistical CSI. We next formulate a joint optimization problem to maximize the sum SE of the system. We first introduce a novel successive Quadratic Transform (successive-QT) algorithm to optimize the transmit power coefficients using the concept of block optimization along with quadratic transform and then use the particle swarm optimization technique to design the RIS phase shifts. Note that most of the existing works on RIS-aided cell-free systems are specific instances of the general scenario studied in this work. We numerically show that i) the RIS-assisted link is more advantageous at lower transmit power regions where the direct link between AP and user is weak, ii) NOMA outperforms orthogonal multiple access schemes in terms of SE, and iii) the proposed joint optimization framework significantly improves the sum SE of the system.  
\end{abstract}   

\vspace{-2pt}
\IEEEpeerreviewmaketitle
\vspace{-22pt}
\section{Introduction} 
Fifth-generation (5G) cellular communication uses massive multiple-input multiple-output (mMIMO) technology to enable precise beamforming toward any location in the cell. With the commercial success of 5G deployments, researchers have already started to look beyond with an aim to provide a better user experience with widespread coverage, high spectral efficiency (SE), and low latency. Many new technologies have emerged, such as cell-free mMIMO~\cite{NgoAYLM17}, reconfigurable intelligent surfaces~\cite{RenzoZDAYRT20}, non-orthogonal multiple access~\cite{NOMA_Base_Paper}, among others, as a probable candidate for the beyond 5G communication systems. 

Cell-free mMIMO system is a distributed architecture where a large number of geographically distributed access points (APs) jointly serve a number of users randomly distributed over a large area. Cell-free networks achieve a higher and uniform signal-to-noise ratio (SNR) within the coverage area than conventional cellular networks~\cite{NayebiAMYR17}. Many aspects of cell-free systems are being actively researched, with a focus on technical foundations~\cite{NgoAYLM17},  resource allocation~\cite{NayebiAMYR17}, signal processing~\cite{ChienBL20}, practical implementation~\cite{MasoumiE20, WangZDZJY20}, among others.
However, under harsh propagation conditions, cell-free systems cannot guarantee a good quality-of-service for all users. In addition, the dense deployments of APs may result in a significant increase in the overall cost.

Reconfigurable intelligent surface (RIS), which is composed of passive elements, has emerged as a promising cost-effective technique by generating favorable propagation conditions between the APs and users \cite{ZhangDZLXZLS21}. 
The existing literature on RIS focuses on different aspects, for example, theory and design of RIS~\cite{DBLP:IRS_5}, performance analysis~\cite{ZhangZ20a}, applications and optimization of RIS-assisted wireless networks \cite{ZhangDZLXZLS21}, among others. {Reference~\cite{Q2} investigates an RIS assisted downlink system and derived fundamental capacity limits considering dirty paper coding (DPC) scheme. The authors in~\cite{Q2} showed that the sum-rate advantage of RIS-aided DPC over RIS-aided zero-forcing diminishes as the number of RIS elements grows significantly.} There are few papers which integrate RIS in cell-free systems to further improve the system performance~\cite{ZhangD21, abs-2104-08648, 42}. In RIS-assisted cell-free systems, the RIS is deployed around the APs and users to create favorable propagation conditions through low-cost reconfigurable reflections. 
Zhang \textit{et al.} in \cite{ZhangD21} considered a downlink RIS-assisted cell-free MIMO system and assumed perfect channel state information (CSI) and uncorrelated channels.  
Chien \textit{et al.} in~\cite{abs-2104-08648} derived the closed-form and asymptotic SE expressions for a RIS-assisted cell-free mMIMO system over spatially correlated channels. Noh \textit{et al.} in \cite{42} proposed a new and efficient two-timescale algorithm to maximize the minimum achievable rate of a cell-free MIMO system powered~by~RIS.

\begin{figure*}       
    \includegraphics[scale=0.395]{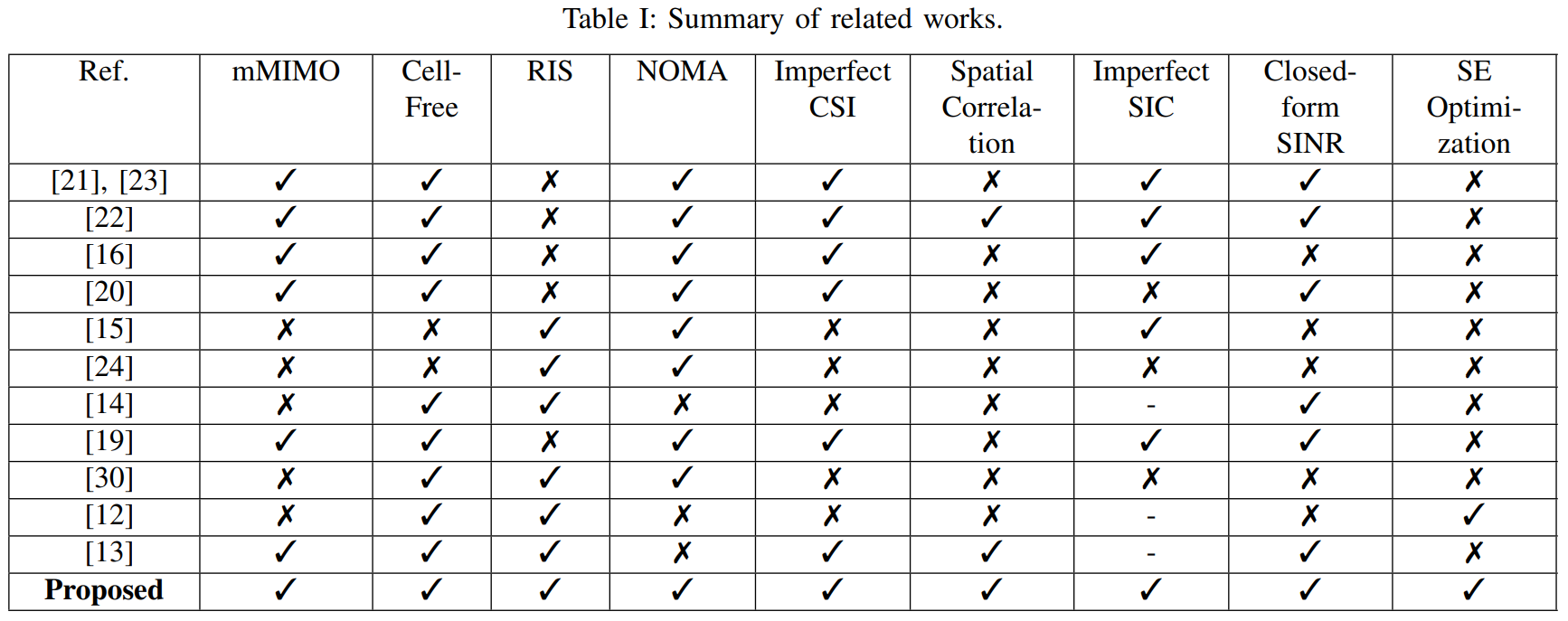} 
    \label{Table-I} \\[-25pt]
\end{figure*}
Non-orthogonal multiple-access (NOMA) serves multiple users on the same orthogonal time/frequency resource, thereby allowing a large number of users to be connected to the network~\cite{ChengLLTP21}. Incorporating NOMA into a cell-free system results in higher throughput, massive connectivity with low latency, and high reliability with user fairness~\cite{34}.  
Existing literature on MIMO NOMA has shown several advantages of using NOMA over orthogonal multiple access (OMA) in different aspects~\cite{SenaCFNCDDPS20,ChengLLTK21}. The works in~\cite{R1, 34,Perfect_SIC,BasharCBNHX20,32,33} investigated the NOMA-assisted cell-free mMIMO system. {The authors in~\cite{R1} compared the sum-rate of NOMA-aided cell-free systems against its OMA counterpart and showed that NOMA achieves higher sum rate~gain~over OMA when the number of simultaneously served users grows~large.} Reference~\cite{34} compared the achievable rate performance of NOMA and OMA in the downlink transmission {assuming imperfect successive interference cancellation (SIC)}. Reference~\cite{Perfect_SIC} considered the impact of linear-combination of channel estimations to calculate the downlink rate per user. Reference~\cite{BasharCBNHX20} derived a closed-form downlink SE expression assuming estimated CSI and imperfect SIC. Zhang \textit{et al.} in~\cite{32} considered a NOMA-based cell-free mMIMO system and derived the closed-form SE expressions with three different estimators. Rezae \textit{et al.} in~\cite{33} derived the downlink achievable sum-rate with three standard precoders for NOMA-aided cell-free mMIMO systems and showed that NOMA can support significantly more users compared to OMA at the same time-frequency resources.  

NOMA and RIS are complementary technologies where NOMA helps in improving the throughput and connectivity of RIS systems, and RISs ensure that the propagation channel is intelligently optimized for the implementation of NOMA. Cheng \textit{et al.} in \cite{ChengLLTP21} considered a RIS-aided NOMA system assuming imperfect SIC. Zheng \textit{et al.} in \cite{39} analyzed the performance of a downlink RIS-assisted MIMO-NOMA system assuming perfect SIC.
{Reference~\cite{TD3} considered a downlink RIS assisted MIMO NOMA system and analyzed the capacity region of RIS-assisted NOMA, comparing it with the rate region of OMA from an information-theoretic perspective. Their findings demonstrate that NOMA can match the performance of DPC. In~\cite{Q3}, the authors mentioned that by employing RIS, NOMA can achieve same performance as DPC but with reduced computational complexity. Reference \cite{N1,N2} showed that RIS-aided NOMA performs better in minimizing the total transmit power compared to the space
division multiple access (SDMA) schemes. The authors in~\cite{BasharCBNHX20} considered a NOMA assisted cell-free mMIMO system and proposed OMA/NOMA mode switching scheme for maximizing the average per-user bandwidth efficiency of the system that depends on both the channel coherence time and the total number of users. Reference~\cite{N3} presented a novel framework for the long-term control and deployment design of RIS-enhanced MISO-NOMA systems. The authors demonstrated that utilizing NOMA in RIS systems can lead to higher energy efficiency compared to those employing the OMA scheme.}

We see from Table I that the existing literature has considered i) RIS cell-free MIMO~\cite{ZhangD21, abs-2104-08648, 42}, ii) RIS NOMA~\cite{ChengLLTP21,39} and iii) Cell-free NOMA~\cite{R1,34,Perfect_SIC,BasharCBNHX20,32,33}. However,  the combination of cell-free, RIS, and NOMA has not been explored in the earlier literature, except recently in \cite{R2}. {Reference~\cite{R2} considered a RIS assisted uplink cell-free NOMA system assuming perfect CSI/SIC and uncorrelated channel, with an aim to reduce the energy consumption with a proposed distributed RIS-enhanced architecture.} In addition, the literature on cell-free RIS except~\cite{abs-2104-08648} has considered perfect CSI and uncorrelated channels.  Recall the authors in~\cite{abs-2104-08648} investigated {cell-free RIS OMA system} over spatially correlated channels and analyzed SE considering random RIS phases and equal transmit power coefficients.  The current work plugs these gaps. Through this work, we envision a wireless system where both RIS and NOMA are integrated into a cell-free network, which reaps all the advantages of these technologies, resulting in a better SE. \textit{This is the first work which considers NOMA into a RIS-assisted cell-free massive MIMO system,  considering imperfect CSI, imperfect SIC and spatial correlation at the RIS.}
The \textbf{main contributions} of this paper are as~follows: 

\begin{itemize} \addtolength{\leftskip}{-10pt}
    \item We consider a RIS-assisted cell-free mMIMO NOMA system and take into account i) {imperfect instantaneous CSI}, ii) imperfect SIC, and iii) spatial correlation among the elements of the RIS, and derive a closed-form expression for downlink SE using the statistical estimate of the channel and conjugate beamforming.  We consider a more feasible and practical scenario.  
The derived SE expression is general and simplifies to the existing works \cite{NgoAYLM17, Perfect_SIC, BasharCBNHX20, abs-2104-08648} through appropriate modifications.   
    
    \item We propose a novel alternating optimization framework to maximize the sum SE by optimally allocating the power control coefficients from each AP to user and by designing an optimal RIS phase matrix. We use the derived closed-form SE expression, which depends on large-scale channel statistics, to formulate the SE maximization problem. The SE is a non-concave function of the transmit power control coefficients and the RIS phase shifts. We first maximize it by optimizing the transmit power control coefficients by developing a low-complexity successive quadratic transform (QT) based algorithm, where we introduce the block optimization concept along with the QT technique. We next optimize the RIS phase shifts using the swarm-based particle swarm optimization (PSO) algorithm. We finally propose a combined framework to jointly optimize the power control coefficients and RIS phase shifts. 
    
    \item Numerical investigations reveal that NOMA outperforms OMA in terms of SE for the RIS-assisted cell-free system, and the RIS-assisted link is more advantageous in lower transmit power regions where the direct link between AP and user is weak. 
    We show the effectiveness of the proposed optimizations through the following key findings: i) optimizing transmit power control coefficients is more effective at the higher transmit power region whereas optimizing the RIS phase shifts is more effective in the lower transmit power region, and ii) with optimization the impact of the intra-cluster interference on the SE significantly reduces especially, at the higher transmit power region. 
\end{itemize} 
\begin{figure*} 
    \includegraphics[scale=0.25]{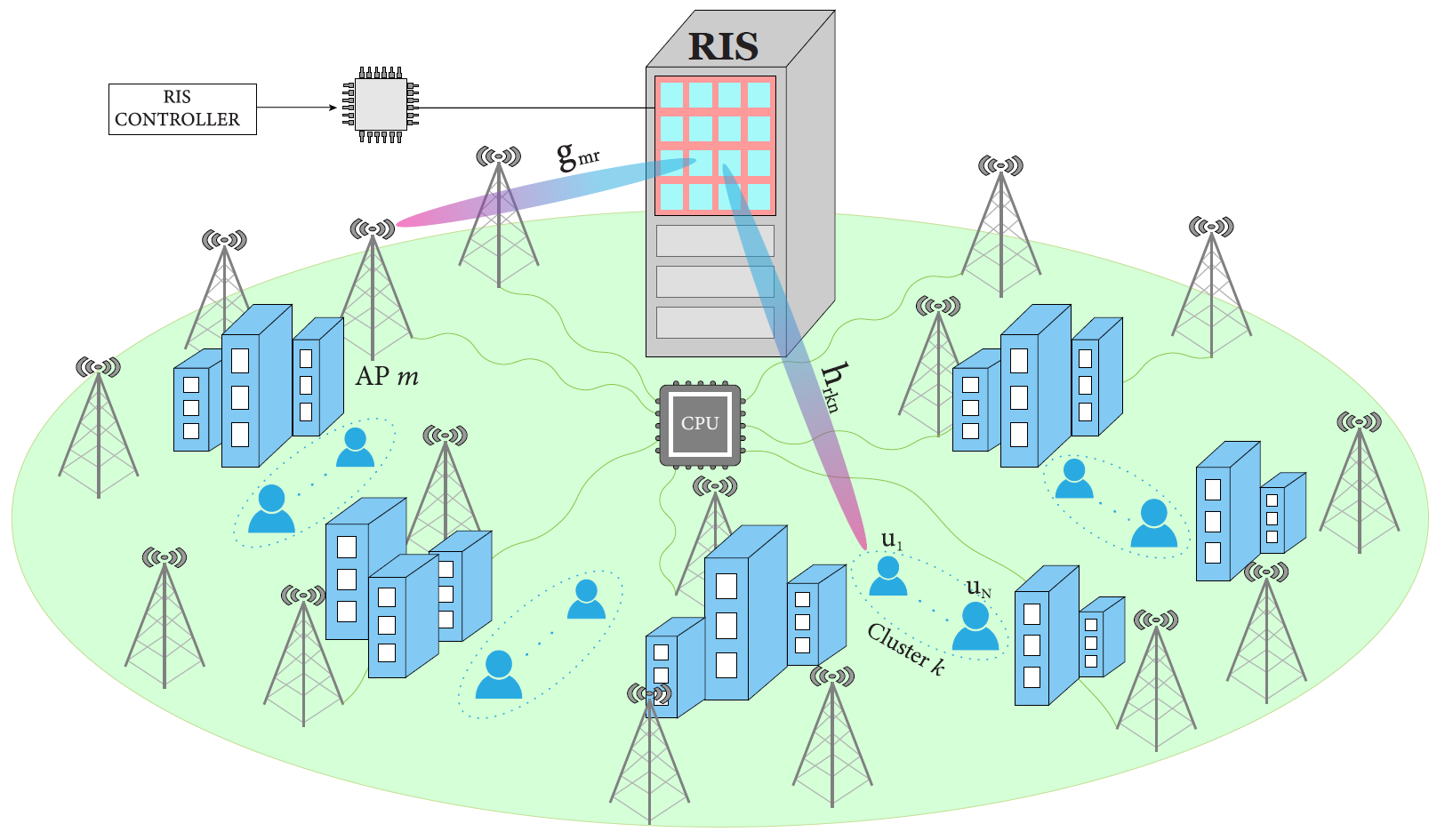}
    \caption{\small RIS-assisted cell-free mMIMO NOMA system. \\[-10pt]} 
    \label{fig:system_model} 
\end{figure*} 

\vspace{-5pt}
\section{System Model} 
We consider a RIS-assisted cell-free mMIMO NOMA system, as shown in Fig. \ref{fig:system_model}, wherein $M$ APs communicate with a total number of $KN$ single-antenna users. We introduce the RIS\footnote{The extension to this work for multiple RISs can be taken as future work.} to the system that consists of $L$ number of passive reflecting elements\footnote{{By leveraging the active RIS in the current system model which allows for smart control of the amplification gain at each RIS element, the active RIS may perform better than the passive RIS~\cite{AR1,AR2}.  Future work can consider modifications to the proposed framework by incorporating active RIS over passive RIS.}}.  APs, RIS, and users are randomly deployed in a large geographical area. The APs are connected to a central processing unit through optical fronthaul links~\cite{NgoAYLM17}. We consider that both direct and reflected links (via RIS) exist  between APs and users, thereby providing more diverse transmission. The APs communicate with $K$ clusters having $N$ single-antenna users each. The APs serve all the users on a particular cluster over the same spectral resources by employing NOMA. We assume a weak direct path scenario to highlight the advantages obtained from RIS where the obstacles block the channels between APs and users. {In addition, the RIS is installed at a certain height such that it is not affected by obstacles.} All the nodes in the network operate in time-division duplex mode. Let us assume a coherence interval of $\tau_c = \tau_p+\tau_d$ symbols, where $\tau_p$ denotes the duration of uplink training and $\tau_d$ denotes the duration of downlink data transmission. 
{Note that, the RIS phases in the data transmission and channel estimation  are the same~\cite{abs-2104-08648}. Under the assumption of perfect statistical channel knowledge, the CPU can first run the optimization in Section IV to determine the RIS phases.  Then, uplink training (for instantaneous CSI estimation in Section II-B) and downlink data transmission (Section II-C) are implemented based on the determined RIS phases.} In the following subsections, we discuss the channel modeling followed by uplink training and downlink~data transmission. 

\vspace{-10pt}
\subsection{Channel Model} \vspace{-2pt}
The channel from $m$th AP to RIS and RIS to $n$th user in the $k$th cluster,  denoted by $\mathbf{g}_{mr} \in \mathbb{C}^{{L \times 1}}$ and  $\mathbf{h}^{H}_{rkn} \in \mathbb{C}^{1 \times L}$ respectively are modelled as
\begin{equation}
\mathbf{g}_{mr} = \mathbf{{R}}_{mr}^{1/2} \mathbf{\bar{g}}_{mr}  \quad \text{and} \quad
    {\mathbf{h}_{rkn}} = {\mathbf{{R}}_{rkn}^{1/2}} \mathbf{\bar{h}}_{rkn},
\end{equation}
where the matrices $\mathbf{R}_{mr} \in \mathbb{C}^{L \times L}$ and $\mathbf{{R}}_{rkn} \in \mathbb{C}^{L \times L}$ represent the covariance matrix of channel $\mathbf{g}_{mr}$ and $\mathbf{h}_{rkn}$, respectively.  We model these matrices as $\mathbf{R}_{mr} = \beta_{mr} d_H d_V \mathbf{R}$  and  $\mathbf{R}_{rkn} = \beta_{rkn} d_H d_V \mathbf{R}$,  where scalars $\beta_{mr}$ and $\beta_{rkn}$ represents the large scale coefficients of channels from $m$th AP to RIS and RIS to $n$th user in the $k$th cluster, respectively~\cite{abs-2104-08648}. The scalars $d_H$ and $d_V$ represent the horizontal and vertical widths of each RIS element and the matrix $\mathbf{R} \in \mathbb{C}^{L \times L}$ denote the spatial correlation at RIS.  The vectors $\mathbf{\bar{g}}_{mr} \sim \mathcal{CN}\left(0, {\mathbf{{I}}_{L}}\right) $ and $\mathbf{\bar{h}}_{rkn} \sim \mathcal{CN}\left(0, {\mathbf{{I}}_{L}}\right) $ represents the small scale fading.  We denote the direct channel between the $m$th AP and $n$th user in $k$th cluster as $l_{mkn}=\beta_{mkn}^{1/2}\bar{l}_{mkn}$,  where $\beta_{mkn}$ represents the large scale fading coefficient and $\bar{l}_{mkn}\sim \mathcal{CN}\left(0,1\right)$ denotes the small-scale fading.
The aggregate downlink channel between the $m$th AP and the $n$th user in $k$th cluster is~given~as\footnote{{The existence of both direct and reflected links is the nature of signal propagation. There may be some very weak links. The channel estimation is performed via uplink training phases (under time division duplex operation), and hence, the channel estimation overhead depends mainly on the number of users. The estimation of weak links at each APs does not affect the channel estimation overhead much. Importantly, in Section IV, power allocation is done which will implicitly removes the weak links.}}
\begin{equation}\label{virtual_channel}
    {u}_{mkn} = l_{mkn} + \mathbf{h}_{rkn}^H \boldsymbol\Theta_r \mathbf{g}_{mr}.
\end{equation}
Here the phase shift matrix $\boldsymbol\Theta_r = \text{diag} (e^{j\theta_1},\cdots\!,e^{j\theta_{L}})\in \mathbb{C}^{L \times L}$,  where $\theta_i$ for $i=1,\cdots \!,L$ denotes the phase added by the $i$th element of RIS and $\theta_i\sim\mathcal{U}[-\pi,\pi)$~\cite{ZhangDZLXZLS21}. 

\vspace{-10pt}
\subsection{Uplink Training} \vspace{-2pt} 
We consider a pilot transmission strategy where all the $N$ users in the $k$th cluster transmit the same pilot signal of length $\tau_p\geq K$ to the APs. 
Let $\sqrt{\tau_p}\mathbf{\Phi}_{k} \in \mathbb{C}^{\tau_p \times 1}$ denotes pilot transmitted by the $n$th user in the $k$th cluster. The pilot satisfies i) $\|\mathbf{\Phi}_k\|^2 = 1$, ii)~$\mathbf{\Phi}_k^H \mathbf{\Phi}_{k'} = 0 $ for $k \ne k'$. The pilot signal received by the $m$th AP,  denoted as $\mathbf{y}_{pm} \in \mathbb{C}^{\tau_p\times 1}$,  can be expressed as 
\begin{equation}\label{y_pm}
\mathbf{y}_{pm} = \sum\limits_{k=1}^{K} \sum\limits_{n=1}^{N}\sqrt{\tau_p \rho_p} {u}_{mkn} \boldsymbol{\Phi}_{k} + \mathbf{w}_{pm}.
\end{equation}
Here the scalar $\rho_p$ denotes the normalized transmit signal-to-noise-ratio (SNR) of the pilot signal and $\mathbf{w}_{pm}\sim\mathcal{CN}(0,\mathbf{I}_{\tau_p})$ denotes the additive white Gaussian noise (AWGN) at the $m$th AP.  The $m$th AP estimates the channel by projecting $\mathbf{\Phi}_k$ onto  the received signal $\mathbf{y}_{pm}$,  and the resultant signal $\breve{{y}}_{pmk}$ is written as follows
\begin{equation}\label{y_pmk}
\breve{{y}}_{pmk} = \boldsymbol{\Phi}_{k}^H \mathbf{y}_{pm} = \sqrt{\tau_p \rho_p} \sum\limits_{n=1}^{N}{u}_{mkn} + \mathbf{\Phi}_k^H \mathbf{w}_{pm},
\end{equation}
where $\mathbf{\Phi}_k^H \mathbf{w}_{pm} \sim \mathcal{CN}(0,1)$.  
We, similar to~\cite{BasharCBNHX20}, estimate the linear combination of all the users channels in the same cluster i.e., $z_{mk} = \sum_{n=1}^N u_{mkn}.$ 
The following key result in Proposition~\ref{Proposition-1} is helpful in computing the channel estimate.  
\vspace{-2pt}
\begin{Proposition} \label{Proposition-1}
For a RIS-assisted cell-free mMIMO NOMA system,  the linear minimum mean square error (LMMSE)\footnote{{The effective channel $z_{mk}$ is the sum of many random variables, and hence, its distribution is very close to the Gaussian distribution (due to the central limit theorem). As a result, using LMMSE estimation is nearly optimal.}} estimate of $z_{mk}$ based on \eqref{y_pmk} is given as~\cite{SIG-093}
\begin{equation}\label{channel_estimation}
    {\hat{z}}_{mk} = c_{mk}\left( \sqrt{\tau_p \rho_p} \sum_{n=1}^N u_{mkn} + \mathbf{\Phi}_k^H \mathbf{w}_{pm} \right),
\end{equation}
where $c_{mk}$ is given as
\begin{equation}\label{c_mk}
    c_{mk} = \frac{\sqrt{\tau_p \rho_p } \sum_{n'=1}^{N}\delta_{mkn'}}{{\tau_p \rho_p }  \sum_{n'=1}^{N} \delta_{mkn'} + 1}.
\end{equation} 
Here the constant $\delta_{mkn} =  \beta_{mkn} + \text{Tr}\left(\mathbf\Theta_r \mathbf{R}_{mr} \mathbf\Theta_r^H \mathbf{R}_{rkn} \right)$ represents the variance of combined channel which includes both direct and indirect channels.
\end{Proposition}

\vspace{-15pt}
\subsection {Downlink Data Transmission Phase} \vspace{-2pt}
Each AP uses the channel estimates in~\eqref{channel_estimation} and designs conjugate beamforming to precode the desired symbols.  The $m$th AP transmits a precoded signal $x_{m} = \sum_{k=1}^{K}   {\hat{z}}_{mk}^{*} q_{k}$ to all the users, where $q_{k} = \sum_{n = 1}^{N} \sqrt{\eta_{mkn}\rho_d} s_{kn}$ is the NOMA signal for the $k$th cluster. Here $\rho_d$ denotes the normalized SNR in the downlink and $s_{kn}$ is the message signal for the $n$th user associated with the $k$th cluster, which satisfies $\mathbb{E}\{|s_{kn}|^2\} = 1$. The scalar $\eta_{mkn}$ denotes the power control coefficient,  which is designed such that it satisfies the power constraint at each AP, i.e.,  $\mathbb{E}\{|x_{m}|^2\} \leq \rho_d$ which leads to
\vspace{-2pt}
\begin{equation} \label{power_constraint}
\sum_{k=1}^{K} {\sum_{n=1}^{N}} \gamma_{mk} \eta_{mkn} \leq 1,
\end{equation} 
where 
\begin{equation} \label{gamma_mk}
 \gamma_{mk} = \mathbb{E}\left\{ \lVert{\hat{z}_{mk}}\rVert^2 \right\} = \sqrt{\tau_p \rho_p } c_{mk}\sum_{n'=1}^N \delta_{mkn'}.  
\end{equation}

The signal received by the $n$th user in the $k$th cluster is 
\begin{align}\label{rxsignal}
r_{kn} &= \sum\limits_{m = 1}^M  {u}_{mkn} x_{m} + {v}_{kn},\nonumber \\ 
&= \sum\limits_{m = 1}^M \sum_{k^\prime=1}^{K} \sum\limits_{n^\prime = 1}^{N} \sqrt{\eta_{m k^\prime n^\prime} \rho_d} u_{mkn} \hat{z}_{m k^\prime}^{*}  s_{k^\prime n^\prime} + v_{kn},
\end{align}
where $v_{kn} \sim \mathcal{CN}(0,1)$ denotes the additive noise at the $n$th user in the $k$th cluster. The received user signal $r_{kn}$ is next re-written in~\eqref{rx_components} (\textit{shown at the top of the next page}) to indicate the i) desired signal, ii) intra-cluster interference, iii) inter-cluster interference, and iv) additive noise.  
\begin{figure*}
   \begin{align}\label{rx_components}
    {r}_{kn} =\underbrace{\sum\limits_{m = 1}^M \sqrt{\eta_{mkn} \rho_d} {u}_{mkn} {\hat{z}}_{mk}^{*}}_{\text{Desired signal}} s_{kn} +\! \underbrace{\sum\limits_{m = 1}^M  \sum\limits_{n^\prime \neq n}^{N} \sqrt{\eta_{m k n^\prime} \rho_d}{u}_{mkn} {\hat{z}}_{m k}^{*} }_{\text{Intra-cluster interference}}s_{kn^\prime} +\! \underbrace{\sum\limits_{m = 1}^M  \sum_{k^\prime \neq k}^{K} \sum\limits_{n^\prime = 1}^{N} \sqrt{\eta_{m k^\prime n^\prime} \rho_d}{u}_{mkn} {\hat{z}}_{m k^\prime}^{*} }_{\text{Inter-cluster interference}}s_{k^\prime n^\prime} +\!\!\! \underbrace{\quad {v}_{kn} \quad }_{\text{Additive noise}}\!\!, \nonumber\\[-10pt]
    \end{align}  
\vspace{-25pt}
\end{figure*}
The users associated with the $k$th cluster mitigate the intra-cluster interference by performing SIC. Keeping~\eqref{Gamma_cf} in mind\footnote{{To enable SIC in every NOMA cluster, the users are ordered based on the overall effective channel gain.  We, therefore, use the derived closed-form SE expression in~\eqref{Gamma_cf} to obtain the effective channel gain.}}, we define the virtual channel for $n$th user in the $k$th cluster as
\begin{align}\label{user_ordering} 
\mathbf{u}_{kn} = \left[ \frac{\gamma_{1k} \delta_{1kn}}{\sum_{p=1}^N \delta_{1kp}}, \frac{\gamma_{2k} \delta_{2kn}}{\sum_{p=1}^N \delta_{2kp}}, \cdots\!,  \frac{\gamma_{Mk} \delta_{Mkn}}{\sum_{p=1}^N \delta_{Mkp}}\right]^T. 
\end{align} 
To enable SIC,  we assume that in every NOMA cluster, the users are ordered based on the virtual channel information as
$\lVert{\mathbf{u}_{k1}}\rVert \geq \lVert{\mathbf{u}_{k2}}\rVert \geq \cdots \geq\lVert{\mathbf{u}_{kN}}\rVert$.
The $n$th user using SIC first cancels the intra-cluster interference from $ n<n'\leq N$ users.  To enable this, the user employs the statistical estimate $\mathbb{E}\left\{ \sum_{m=1}^M u_{mkn} \hat{z}_{mk}^* \right\}$,  considering the fact that the effective channel hardens for large number of APs~\cite{SIG-093}. The scalar statistical estimate remains unchanged for several coherence intervals and, therefore, is assumed to be known~\cite{SIG-093}.  After performing SIC,  the user decodes its own signal by treating the signal from the first ($n-1$) users as inherent intra-cluster interference. Hence, the received signal \eqref{rx_components} after an imperfect SIC process, $r_{kn}^{\text{ISIC}} = r_{kn} - \sum_{i=n+1}^{N} \;\mathbb{E}\left\{ \sum_{m=1}^M \sqrt{ \eta_{mki}{\rho_d}} u_{mkn} \hat{z}_{mk}^* \right\} s_{ki}$ can be expressed as \eqref{rx_components_sic} (\textit{shown at the top of the next page}),  
\begin{figure*}
    \begin{flalign}\label{rx_components_sic}
    {r}_{kn}^{\text{ISIC}} = &\underbrace{\sum\limits_{m = 1}^M \sqrt{\eta_{mkn} \rho_d}{u}_{mkn} {\hat{z}}_{m k}^{*} }_{\mathcal{T}_0:\;\text{Desired signal}} s_{kn} + \underbrace{\sum\limits_{m = 1}^M  \sum\limits_{n^\prime \neq n}^{n-1} \sqrt{\eta_{m k n^\prime} \rho_d}{u}_{mkn} {\hat{z}}_{m k}^{*} }_{\mathcal{T}_1:\;\text{Inherent intra-cluster interference}}s_{kn^\prime} \nonumber + \underbrace{\sum\limits_{m = 1}^M  \sum_{k^\prime \neq k}^{K} \sum\limits_{n^\prime = 1}^{N} \sqrt{\eta_{m k^\prime n^\prime} \rho_d}{u}_{mkn} {\hat{z}}_{m k^\prime}^{*}}_{\mathcal{T}_2:\;\text{Inter-cluster interference}}s_{k^\prime n^\prime} \nonumber\\ & + \underbrace{\sum_{i=n+1}^{N}\sqrt{\rho_d} \bigg(\sum\limits_{m = 1}^M   \sqrt{\eta_{mki}} {u}_{mkn} {\hat{z}}_{m k}^{*} -\mathbb{E}\left\{ \sum_{m=1}^M \sqrt{\eta_{mki}} u_{mkn} \hat{z}_{mk}^* \right\}\bigg) }_{\mathcal{T}_3:\;\text{Residual intra-cluster interference post SIC}}s_{ki}  + \underbrace{ {v}_{kn} }_{\text{Additive noise}}. 
\end{flalign}  
\hrule
\vspace{-15pt}
\end{figure*} 
The received signal after SIC in \eqref{rx_components_sic} depends on the instantaneous channel realizations that are unknown at the users. 

\vspace{-5pt}
\begin{Remark}
{Typically, a wireless system has a diverse set of users distributed randomly within its coverage area.  Within this setup,  there are users with both strong and weak links to the APs. Weak links can result from factors like extended distances or significant obstructions causing large path loss and heavy shadowing.  In our work, we primarily focus on scenarios where we use RIS to enhance the performance of users with weak connections, with the aim of enhancing the overall coverage and efficiency of the whole system.}
\end{Remark}

\vspace{-7pt}
\section{Achievable Downlink Spectral Efficiency}
In this section, we first calculate the ergodic sum SE and then derive a closed-form SE with estimated CSI and imperfect SIC.  The derived closed-form expression depends only on large-scale fading coefficients, which vary only after a certain number of coherence intervals,  and therefore allows a system designer to calculate SE  without performing tedious system-level simulations.  
The ergodic SE of the $n$th user in the $k$th cluster,  can be written using~\eqref{rx_components_sic} as follows
\begin{equation}\label{ergodicsumrate}
\text{SE}_{kn} = \left(1 - \frac{\tau_p}{\tau_c}\right)~ \mathbb{E}\left\{\log_2(1 + \Lambda_{kn})\right\},
\end{equation} 
\begin{equation}\label{Lambdaexpression}
   \!\!\!\text{with}\;\Lambda_{kn} = \frac{P_{kn}^{d}}{ \sum\limits_{n'= 1}^{n-1} \text{I}^{I}_{kn'}+ \sum\limits_{n'= n+1}^{N}\text{I}^{R}_{kn'} + \!\sum\limits_{k'\ne k}^{K}\sum\limits_{n'=1}^{N} \text{I}^{C}_{k'n'} +\! 1},\!\!\!
\end{equation} 
where $\displaystyle{P_{kn}^{d} =  \left| \sum\limits_{m = 1}^M \sqrt{\eta_{mkn} \rho_d} {u}_{mkn} {\hat{z}}_{m k}^{*} \right|^2}$,\\ $\phantom{x}\hspace{21pt}\text{I}^{I}_{kn'} =\left| \sum\limits_{m = 1}^M \sqrt{\eta_{m k n'} \rho_d} {u}_{mkn} {\hat{z}}_{mk}^{*}  \right|^2$,\\
$\phantom{x}\hspace{0pt}\text{I}^{R}_{kn'}={\rho_{d}}\left|\sum\limits_{m = 1}^M \!\!\!\sqrt{\eta_{mkn'}} u_{mkn} \hat{z}_{mk}^*\!\! -\mathbb{E}\!\left\{\!\sum\limits_{m=1}^M \!\!\sqrt{\eta_{mkn'}}u_{mkn} \hat{z}_{mk}^{*}\!\! \right\}\!\right|^2\!$\!,\\
$\phantom{x}\hspace{0pt}\text{I}^{C}_{k'n'} = \rho_d \left|\sum\limits_{m = 1}^M \sqrt{\eta_{mk'n'} } {u}_{mkn} {\hat{z}}_{mk'}^{*} \right|^2$.\\ 
 
The SE expression in~\eqref{ergodicsumrate} is difficult to simplify due to the expectation outside the logarithm.  Also, recall that the channel is estimated at the APs and not at the users. However, the ergodic SE expression assumes that the instantaneous channel information is known to the users. The expression in~\eqref{ergodicsumrate} therefore provides an upper bound on the achievable SE and is used to validate the derived closed-form expression in Theorem~\ref{Theorem 1}.  To derive the closed-form SE,  we utilize the channel hardening bounding technique \cite{SIG-093},  where the signal received after SIC in \eqref{rx_components_sic} can be written as \eqref{rx_components_cf} (\textit{shown at the top of the next page}).  

\begin{figure*}
   \begin{flalign}\label{rx_components_cf}
    {r}_{kn}^{\text{CF}} &= \underbrace{\mathbb{E}\left\{\sum\limits_{m = 1}^M \sqrt{\eta_{mkn} \rho_d} {u}_{mkn} {\hat{z}}_{m k}^{*}  \right\}}_{\text{Desired Signal}}s_{kn} + \underbrace{ \sqrt{\rho_d}\left(\sum\limits_{m = 1}^M {u}_{mkn} {\hat{z}}_{m k}^{*} \sqrt{\eta_{mkn}} - \mathbb{E}\left\{\sum\limits_{m = 1}^M {u}_{mkn} {\hat{z}}_{m k}^{*} \sqrt{\eta_{mkn} }\right\}\right)}_{\text{Beamforming Uncertainty}}s_{kn} \nonumber \\ &\quad +  \underbrace{\sum\limits_{m = 1}^M  \sum\limits_{n^\prime = 1}^{n - 1} {u}_{mkn} {\hat{z}}_{mk}^{*} \sqrt{\eta_{m k n^\prime} \rho_d} }_{\text{Inherent Intra-cluster interference}}s_{kn'} \nonumber + \!\!\! \underbrace{\sum_{n'=n+1}^{N} \!\!\!\sqrt{\rho_d}\left(\sum\limits_{m = 1}^M \sqrt{\eta_{mkn'}} u_{mkn} \hat{z}_{mk}^* - \sqrt{\eta_{mkn'}}\mathbb{E}\left\{ \sum_{m=1}^M u_{mkn} \hat{z}_{mk}^{*} \right\} \right)}_{\text{Residual Intra-cluster interference post SIC}}\! s_{kn'} \nonumber \\&\quad + \underbrace{\sum\limits_{m = 1}^M  \sum_{k^\prime \neq k}^{K} \sum\limits_{n^\prime = 1}^{N} {u}_{mkn} {\hat{z}}_{m k^\prime}^{*} \sqrt{\eta_{m k^\prime n^\prime} \rho_d} }_{\text{Inter-cluster interference}}s_{k^\prime n^\prime} +\underbrace{\quad {v}_{kn} \quad }_{\text{Additive noise}}. 
    \end{flalign}  
\vspace{-25pt}
\end{figure*}
{In most of the existing literature~\cite{ZhangZ20a,ZhangD21,39},  the optimization problem depends on instantaneous CSI and therefore has to be calculated for every coherence interval. On contrary, we aim to derive the closed-form SE expression that depends on the large scale fading coefficients and is therefore valid for several coherence intervals. The derived closed-form expression is then used to maximize the sum SE of the system by optimally allocating transmit power coefficients and by designing the RIS phase shifts.  Thus, the proposed optimization in Section IV can take place after several coherence intervals, which saves significant overhead.} 
We now derive a closed-form SE expression in the following Theorem~\ref{Theorem 1}. 
\vspace{-5pt}
\begin{theorem}\label{Theorem 1}
For RIS-assisted downlink cell-free mMIMO NOMA system with imperfect SIC and LMMSE-based conjugate beamforming, the closed-form SE of the $n$th user in the $k$th cluster over spatially correlated Rayleigh fading channel, which is valid for an arbitrary number of antennas is given by $ \overline{\text{SE}}_{kn}= \left(1 - \frac{\tau_p}{\tau_c}\right) \log_2 \left( 1 + \Gamma_{kn} \right)$, where 
\begin{equation}\label{Gamma}
    {\Gamma}_{kn} = \frac{\text{DS}_{kn}}{\text{BU}_{kn} + \text{IaCI}^{I}_{kn'} + \text{IaCI}^{R}_{kn'} + \text{ICI}_{k'n'}+ 1}.
\end{equation} 
Here $\text{DS}_{kn}$ is the desired signal power, $\text{BU}_{kn}$ is the beamforming gain uncertainty, $\text{IaCI}^{I}_{kn'}$ denotes inherent intra-cluster interference, $\text{IaCI}^{R}_{kn'}$ denotes residual intra-cluster interference and $\text{ICI}_{k'n'}$ represents the inter-cluster interference, where the following definitions hold:
\begin{align}\label{DS_BU_ICI}
&\text{DS}_{kn} =\left|\mathbb{E}\left\{ \sum\limits_{m = 1}^M \sqrt{\eta_{mkn} \rho_d} {u}_{mkn} {\hat{z}}_{m k}^{*}  \right\}\right|^2, \nonumber \\ 
&\text{BU}_{kn} =\rho_d \mathbb{E}\left|\sum\limits_{m = 1}^M\!\!\sqrt{\eta_{mkn}}\!{u}_{mkn} {\hat{z}}_{m k}^{*}\!-\!\mathbb{E}\left[\sum\limits_{m = 1}^M\!\!\sqrt{\eta_{mkn}} {u}_{mkn} {\hat{z}}_{mk}^{*} \right]\!\right|^2\!\!\!, \nonumber \\ 
&\text{IaCI}^{I}_{kn'} =\sum\limits_{n' = 1}^{n-1} \mathbb{E}\left\{\text{I}^{I}_{kn'}\right\}, \quad \text{IaCI}^{R}_{kn'} =\sum\limits_{n' = n+1}^{N} \mathbb{E}\left\{\text{I}^{R}_{kn'}\right\},\nonumber \\ 
&\text{ICI}_{k'n'}=\sum\limits_{k' \ne k}^{K}\sum\limits_{n' = 1}^{N} \mathbb{E}\left\{\text{I}^{C}_{k'n'}\right\}. 
\end{align} 
After substituting \eqref{DS_appendix}-\eqref{ICI_appendix} into \eqref{Gamma}, the closed-form SINR $\Gamma_{kn}$ can be expressed as \eqref{Gamma_cf} (shown at the top of page~$6$).\vspace{10pt} 
\begin{figure*}
   \begin{flalign}\label{Gamma_cf}
    {\Gamma_{kn}} = \frac{\left| \sum\limits_{m=1}^M {\sqrt {\eta_{mkn}\rho_d} \gamma_{mk} \frac{\delta_{mkn}}{\sum_{i=1}^N \delta_{mki}}  }\right|^2}{\left\{\Large\substack{\sum\limits_{m=1}^M {\eta_{mkn}\rho_d} \delta_{mkn} \gamma_{mk}\;  +\; \sum\limits_{n'= 1}^{n-1}\left( \left|\sum\limits_{m=1}^M \sqrt{\eta_{mkn'}\rho_d} \gamma_{mk} \frac{\delta_{mkn}}{\sum_{i=1}^{N}\delta_{mki}}\right|^2  +\; \sum\limits_{m=1}^M \eta_{mkn'}\rho_d \delta_{mkn}\gamma_{mk} \right)\\ +\; \sum\limits_{n^\prime = n + 1}^{N} \sum\limits_{m=1}^M \eta_{mkn'}\rho_d \delta_{mkn} \gamma_{mk}\; +\; \sum\limits_{k^\prime \neq k}^{K} \sum\limits_{n^\prime = 1}^{N
    }\sum\limits_{m=1}^M \eta_{mk'n'}\rho_d \delta_{mkn} \gamma_{mk'}\; + \;1}\right\}}.
    \end{flalign} 
\vspace{-1pt}
\end{figure*}

\end{theorem} 
\vspace{-15pt}
\begin{proof}
Refer to Appendix \ref{A}.
\end{proof} 
\vspace{-2pt}
\begin{Remark} {{Table II shows that the closed-form SE expression derived in Theorem \ref{Theorem 1} simplifies their counterparts in \cite{NgoAYLM17}, \cite{abs-2104-08648}, \cite{Perfect_SIC}, \cite{BasharCBNHX20} with suitable modifications. Our work, therefore, not only simplifies to the existing works but also fills the gaps as highlighted above in the existing literature.}} 
\end{Remark}

\begin{figure*}       
    \includegraphics[scale=0.39]{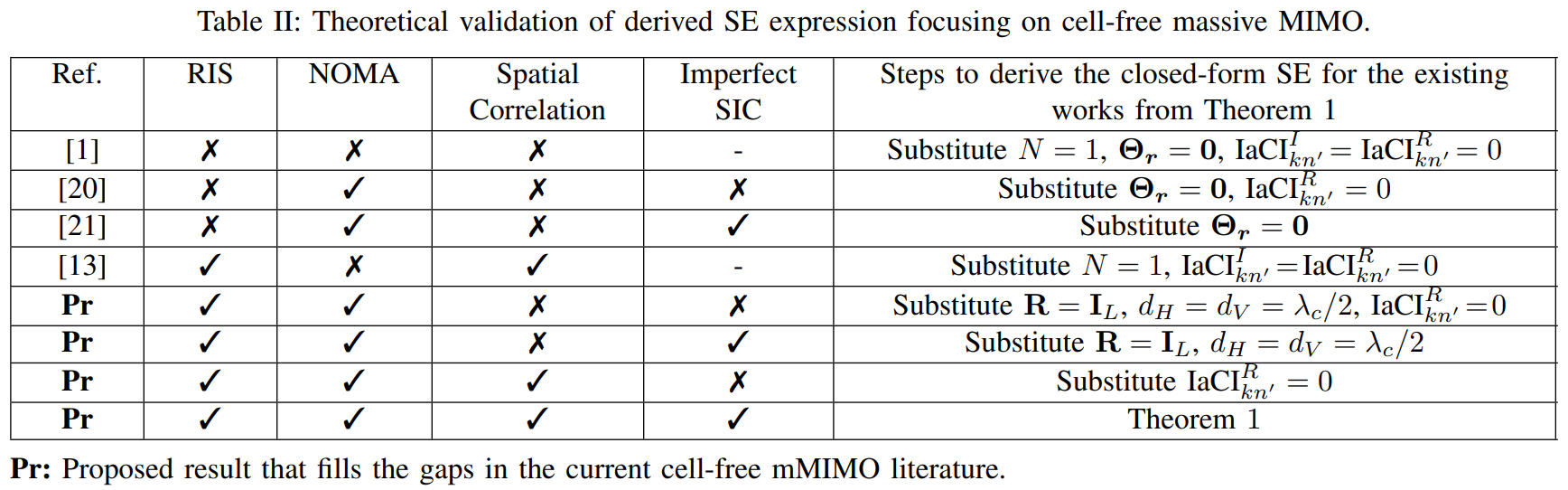} 
    \label{Table-II}   
    \hrule 
\vspace{-15pt}   
\end{figure*} 
\vspace{-15pt}

\subsection{Intuitive Insights from the Derived Closed-Form SINR Expression in Theorem-\ref{Theorem 1}} 
 
\subsubsection{{Effect of imperfect SIC}}
Unlike previous works in cell-free mMIMO NOMA~\cite{Perfect_SIC},\cite{39} which considered perfect SIC, 
{the presence of imperfect SIC introduces residual intra-cluster interference, denoted as $\text{IaCI}^{R}_{kn'}$ (fourth term in the denominator of $\Gamma_{kn}$ in~\eqref{Gamma_cf}).  Moreover, the users can not perform perfect SIC in a single cluster due to the unavailability of instantaneous downlink CSI, intra-cluster pilot contamination, and channel estimation errors. Recall the residual intra-cluster interference from~\eqref{IaCr_appendix}
\begin{align*} 
\text{IaCI}^{R}_{kn'} \;{=} \sum\limits_{n^\prime = n + 1}^{N} \sum\limits_{m=1}^M {{ \eta_{mkn'}}\rho_d} \delta_{mkn} \gamma_{mk}.
\end{align*} 
From (8),  the expression for estimated channel variance is given by 
\begin{align*}
\gamma_{mk} = \sqrt{\tau_p \rho_p } c_{mk}\sum_{n'=1}^N \delta_{mkn'}, 
\end{align*}  
where $\delta_{mkn} = \beta_{mkn} + \text{Tr}\left( \mathbf{\Theta}_{r} \mathbf{R}_{mr} \mathbf{\Theta}_{r}^{H} \mathbf{R}_{rkn}\right)$.
When the RIS link becomes dominant,  the term $\gamma_{mk}$ and $\text{IaCI}^{R}_{kn'}$ becomes stronger.  Therefore,  ignoring the impact of SIC may give exaggerated SE results.} 
\vspace{2pt}
\subsubsection{{Impact of spatial correlation}} 
Recall the aggregated channel variance from \eqref{c_mk}:
$ \delta_{mkn}=\beta_{mkn} + \text{Tr}\left( \mathbf{\Theta}_{r} \mathbf{R}_{mr} \mathbf{\Theta}_{r}^{H} \mathbf{R}_{rkn}\right).$ For spatially uncorrelated channels, the spatial correlation matrix at the RIS, $\mathbf{R}=\mathbf{{I}}_{L}$ and the scalars $d_H=d_V={\lambda_c}/2$, where $\lambda_c$ is the carrier wavelength. Therefore, the covariance matrices from the AP to RIS and RIS to user can be writen as $\mathbf{R}_{mr}=\beta_{mr}({{\lambda_c}^2}/4)\;\mathbf{{I}}_{L}$ and $\mathbf{R}_{rkn}=\beta_{rkn}({{\lambda_c}^2}/4)\;\mathbf{{I}}_{L}$, respectively. This results in $\delta_{mkn} = \beta_{mkn} + \beta_{mr}\beta_{rkn}({{\lambda_c}^4}/16)\;\text{Tr}(\mathbf{{I}}_{L})$. 
We can observe that $\delta_{mkn}$ becomes independent of the RIS phase matrix $\boldsymbol{\Theta}_r$. Hence for an uncorrelated scenario, the SE becomes independent of $\boldsymbol{\Theta}_r$ and relies only on the number of RIS elements $L$. If there is a decrease in spatial correlation among the elements of the RIS, then the combined channel variance $\delta_{mkn}$ weakly depends on the~RIS~phases.

\subsubsection{{Impact of pilot power and pilot length}}
{Recall the estimated channel variance from \eqref{gamma_mk}: 
\begin{align*}
\gamma_{mk} &= \sqrt{\tau_p \rho_p } c_{mk}\sum_{n'=1}^N \delta_{mkn'}\\ &\stackrel{(a)}{=}\frac{{\tau_p \rho_p } \sum_{n'=1}^{N}\delta_{mkn'}}{{\tau_p \rho_p}  \sum_{n'=1}^{N} \delta_{mkn'} + 1}\sum_{n'=1}^N \delta_{mkn'}\; \stackrel{(b)}{\approx}\sum_{n'=1}^N \delta_{mkn'}\;,  
\end{align*} 
where the equality in (a) is obtained by substituting $c_{mk}$ from \eqref{c_mk} and equality in (b) is acquired by assuming that the $\tau_p \rho_p\!>>~\!\!\!1$. 
When the factor $\tau_p\rho_p$ is high, we observe that the estimated channel variance is equivalent to the sum of the individual channel variances. As the channel quality improves, we can achieve almost the same SE with less number of RIS elements at higher transmit power regions. The selection of pilot length $\tau_p$ and pilot power $\tau_p$ will depend on the trade-off between channel estimation overhead factor ($1- {\tau_p}/{\tau_c}$) and channel estimation quality for a fixed coherence interval as shown in Simulation section (Fig.~\ref{fig:13}).}
\subsubsection{{Special cases}} 
\begin{itemize} 
\item{Case 1:} {For a high pilot power and a high transmit power scenario, assuming all the APs are identical and the individual power constraint for each AP is substituted with a collective power constraint ($\eta_{kn}$) that applies to all APs.
Consider a scenario where all the APs are collocated, i.e., $\delta_{mkn}=\delta_{m'kn}\triangleq \delta_{kn}$, then \eqref{Gamma_cf} can be further simplified to
    \begin{flalign} \label{Gamma_cf7}
        \!\!\!\!\!\!{\widetilde{\Gamma}}_{kn} =\!\frac{{M}\eta_{kn}\delta_{kn}}{\!\!\left\{\Large\substack{{\eta_{kn}}  {\sum\limits_{i=1}^{N}\delta_{ki}}\;  +\; \sum\limits_{n'= 1}^{n-1}{\eta_{kn'}}\left( {M}\delta_{kn'} +\; {\sum\limits_{i=1}^{N}\delta_{ki}} \right) \\+\!\! \sum\limits_{n^\prime = n + 1}^{N}\!\!\eta_{kn'} {\sum\limits_{i=1}^{N}\delta_{ki}}\; +\! \sum\limits_{k^\prime \neq k}^{K} \sum\limits_{n^\prime = 1}^{N}\!\eta_{k'n'} {\sum\limits_{i=1}^{N}\!\delta_{k'i}}}\!\right\}}.\!\!\!
    \end{flalign}
We can observe from \eqref{Gamma_cf7} that the beamforming uncertainty ($\text{BU}_{kn}$), residual intra-cluster interference ($\text{IaCI}^R_{kn}$) and inter-cluster interference ($\text{ICI}_{kn}$) terms are independent of the number of APs ($M$). However, the inherent intra-cluster interference ($\text{IaCI}^I_{kn}$) strongly depends on $M$, which dominates the rest of the interference terms in the denominator of \eqref{Gamma_cf7}, as $M$ grows large. Due to this reason, there is a trade-off between the number of users ($N$) per cluster and the NOMA system performance, i.e., for $N>2$, the SE of the system decreases drastically (we can observe this from Fig. \ref{fig:se_pmax}).} 

\item{Case 2:} {Considering the scenario where each cluster contains a single user only, all the $K$ users are allocated mutually orthogonal pilots, and both pilot power and transmit power are sufficiently high, then \eqref{Gamma_cf} can be reduced to 
\begin{align} \label{Gamma_cf8}
    {\Gamma_{k}} = \frac{\left|\sum\limits_{m=1}^M {\sqrt{\eta_{mk}} \delta_{mk} }\right|^2}{\left\{\Large\substack{ \sum\limits_{m=1}^M {{\eta_{mk}}\delta_{mk}^2}\;+\; \sum\limits_{k^\prime \neq k}^{K} \sum\limits_{m=1}^M \eta_{mk'}\delta_{mk}\delta_{mk'}}\right\}}\!. 
\end{align}
We observe that the numerator of ${\Gamma_{k}}$ scales as $M^2$, whereas its denominator scales as $M$. Thus, in contrast to \eqref{Gamma_cf7}, as $M\to\infty$, the interference components in \eqref{Gamma_cf8} diminish. This is because intra-cluster pilot contamination is eliminated in the analyzed scenario.}
\end{itemize} 

\vspace{-5pt}
\section{Spectral Efficiency Maximization with joint power and phase optimization}\vspace{-2pt}
In this section, we discuss the problem formulation and proposed solution to maximize the sum SE of the RIS-assisted cell-free mMIMO NOMA system. We aim to maximize the sum SE by optimally allocating the downlink power control coefficients $\boldsymbol{\eta}$, and by designing the RIS phase matrix $\boldsymbol{\Theta}_r$. The matrix $\boldsymbol{\eta} \triangleq (\boldsymbol{\eta}_1, \dots, \boldsymbol{\eta}_m, \dots, \boldsymbol{\eta}_M)\in \mathbb{R}^{KN \times M}$, where  $\boldsymbol{\eta}_m\in \mathbb{R}^{KN \times 1}$ includes all the power control coefficients for the $m$th AP. 
In the downlink data transmission phase, the achievable sum SE can be written as
 \begin{equation}\label{SE_bar}
     \overline{\text{SE}} =\!\! \sum\limits_{k=1}^{K}\sum\limits_{n=1}^{N} \overline{\text{SE}}_{kn} \!=\! \left(1-\frac{\tau_p}{\tau_c}\right)\! \sum\limits_{k=1}^{K}\sum\limits_{n=1}^{N}\log_2\!\left(1+\Gamma_{kn}\right)\!, 
 \end{equation}  
 where $\Gamma_{kn}$ is the closed-form SINR of the $n$th user in the $k$th cluster derived in Theorem \ref{Theorem 1}. We see that the sum SE in \eqref{SE_bar} is a function of the downlink transmit power coefficients $\boldsymbol{{\eta}}$ and RIS phase matrix $\boldsymbol{\Theta}_r$. 
{We now cast an optimization Problem \textbf{P1} to maximize the sum SE by jointly optimizing $\boldsymbol{{\eta}}$ and $\boldsymbol{\Theta}_r$ as follows:} \vspace{-5pt}
 \begin{subequations}\label{P1}
 \begin{align} \textbf{P1:}\;\;
     \underset{\boldsymbol{{\eta}},\;\boldsymbol{\Theta}_r}{\text{maximize}} & \left(1-\frac{\tau_p}{\tau_c}\right)\! \sum\limits_{k=1}^{K}\sum\limits_{n=1}^{N}\log_2\left(1+\Gamma_{kn}(\boldsymbol{{\eta}},\boldsymbol{\Theta}_r)\right)\label{P1_a} \\[-15pt] 
     \text{subject to}\; &\sum\limits_{k=1}^{K}\sum\limits_{n=1}^{N} \gamma_{mk}\eta_{mkn}\leq 1,\label{P1_b} \\  
     &\eta_{mkn}\geq 0,\label{P1_c}\\
     &\log_2\left(1+\Gamma_{kn} (\boldsymbol{{\eta}},\boldsymbol{\Theta}_r)\right) \geq R_{\text{min},kn},\label{P1_d}\\  
     &\sum\limits_{m=1}^{M}\sum\limits_{i=1}^{n-1}\!\eta_{mki}\leq\!\!\sum\limits_{m=1}^{M}\!\eta_{mkn}, \;\forall k, \forall n\neq 1,\label{P1_e}\\  
     &\theta_{l} \in [0,2\pi], \quad l\in \{1,2,\dots,L\}.\label{P1_f}
 \end{align} 
 \end{subequations} 
The first constraint in \eqref{P1_b} ensures the maximum limit of the transmit power budget available at each AP, while the second constraint in \eqref{P1_c} imposes non-negativity on the transmit power coefficients. The QoS constraint in \eqref{P1_d} guarantees a minimum data rate per user, where $R_{\text{min},kn}$ denotes the threshold for the $n$th user in the $k$th cluster to achieve the required minimum data rate \cite{Ref.1}. Constraint \eqref{P1_e} imposes a power constraint on the SIC process to ensure its successful execution during transmission \cite{Ref.9}. The constraint in \eqref{P1_f} specifies the range of the phase shifts applied by each~RIS~element. 

Problem \textbf{P1} in (\ref{P1}) is non-concave, and the presence of $\boldsymbol{\Theta}_r$ makes it difficult to solve using standard optimization techniques. Moreover, a major challenge in solving this problem lies in the joint optimization of both $\boldsymbol{\eta}$ and $\boldsymbol{\Theta}_r$. Here, we employ an alternative optimization approach that iteratively solves for $\boldsymbol{\eta}$ and $\boldsymbol{\Theta}_r$. Through this iterative process of solving for $\boldsymbol{\eta}$ and $\boldsymbol{\Theta}_r$ while fixing the other, the algorithm gradually improves the SE. Therefore, we divide Problem $\textbf{P1}$ into two sub-problems i.e., $\textbf{P1A}$ and $\textbf{P1B}$. For a fixed $\boldsymbol{\Theta}_r$, we can reformulate Problem $\textbf{P1}$ to optimize~$\boldsymbol{{\eta}}$~as
\begin{subequations} \label{P1A}
 \begin{align} \textbf{P1A:}\quad
     \underset{\boldsymbol{{\eta}}}{\text{maximize}}\; \left(1-\frac{\tau_p}{\tau_c}\right) \sum\limits_{k=1}^{K}\sum\limits_{n=1}^{N}\log_2\left(1+\Gamma_{kn}(\boldsymbol{{\eta}},\boldsymbol{\Theta}_r)\right)\!,\nonumber\\[-15pt]\label{P1A_a} 
\end{align} 
\begin{align}
     \text{subject to}\;\; &\sum\limits_{k=1}^{K}\sum\limits_{n=1}^{N} \gamma_{mk}\eta_{mkn}\leq 1,\;\; \eta_{mkn}\geq 0,\label{P1A_b}\\[-2pt]
     &\log_2\left(1+\Gamma_{kn}(\boldsymbol{{\eta}},\boldsymbol{\Theta}_r)\right) \geq R_{\text{min},kn},\label{P1A_c}\\
     &\sum\limits_{m=1}^{M}\sum\limits_{i=1}^{n-1}\eta_{mki}\leq\sum\limits_{m=1}^{M}\eta_{mkn},\quad\forall k, \forall n\neq 1.\label{P1A_d} 
 \end{align}
 \end{subequations}
In the next step, we fix $\boldsymbol{{\eta}}$ and optimize $\boldsymbol{\Theta}_r$ to maximize the sum SE as 
\begin{subequations}\label{P1B}
 \begin{align} \textbf{P1B:}\quad
     \underset{\boldsymbol{\Theta}_r}{\text{maximize}} \;\; & \left(1-\frac{\tau_p}{\tau_c}\right) \sum\limits_{k=1}^{K}\sum\limits_{n=1}^{N}\log_2\left(1+\Gamma_{kn}(\boldsymbol{{\eta}},\boldsymbol{\Theta}_r)\right)\!,\nonumber\\[-10pt]\label{P1B_a} \\
     \text{subject to}\;\;  
     &\theta_{l} \in [0,2\pi], \quad l\in \{1,2,\dots,L\}.\label{P1B_b}
 \end{align} 
 \end{subequations} 
\subsection {Successive QT-based Power Allocation Optimization}  
The objective of Problem $\textbf{P1A}$ contains a ratio term and the constraints are non-linear, leading to a non-convex fractional programming problem that can not be solved in its original form. Due to the non-convexity, we can not directly use the standard fractional programming methods such as Dinkelbach's transform, Charnes-Cooper transform among others to solve $\textbf{P1A}$. Additionally, Problem $\textbf{P1A}$ involves a massive number of optimization variables, i.e., $MKN$, which introduces a substantial level of complexity to the problem. To develop a low-complexity algorithm, we introduce the block optimization \cite{Ref.10} concept along with QT \cite{QT_firstpaper}, which tackles Problem $\textbf{P1A}$ by successively optimizing the transmit powers of each AP while keeping the transmit powers of other APs fixed.  
We introduce the successive-QT optimization, which is described in the following proposition. 

\begin{Proposition} \label{Proposition-2}
Consider a sum-of-functions-of-ratios optimization problem 
\begin{align}\label{Eq:30}
    \underset {\mathbf{Z}} {\text{maximize}} \sum\limits_{i=1}^{\mathcal{K}} f_i\left(\frac{C_i(\mathbf{Z})}{D_i(\mathbf{Z})}\right), \quad \text{subject to}\quad \mathbf{Z}\in \mathcal{Z}, 
\end{align} \\[2pt]
where $f_i(\cdot)$ is a differentiable function, $\mathbf{Z}\in \mathbb{R}^{\mathcal{K} \times \mathcal{M}}$ is the optimization variable, and $\mathcal{Z}$ is a convex set. Both the numerator $C_i(\mathbf{Z})$ and denominator $D_i(\mathbf{Z})$ are functions that map $\mathbb{R}^{n}$ to positive and strictly positive real numbers, respectively for $i=1,\dots, \mathcal{K}$. The variable $\mathbf{Z}$ is divided into $\mathcal{M}$ blocks, denoted as $\mathbf{Z}=(\mathbf{z}_1, \dots, \mathbf{z}_\mathcal{M})$, where each block belongs to a closed convex subset $\mathcal{Z}_{m}$. It should be noted that the block optimization technique divides the original problem into $\mathcal{M}$ subproblems. These subproblems are optimized successively, one at a time, while the remaining variable blocks are held fixed. When problem \eqref{Eq:30} is optimized with respect to the ${m}$th variable block $\mathbf{z}_{m}$, it can be expressed as  
 \begin{align}\label{Eq:31}
     \underset {\mathbf{z}_m} {\text{maximize}} \sum\limits_{i=1}^{\mathcal{K}} f_i\left(\frac{C_i(\mathbf{z}_m, \mathbf{z}_{m}^\mathsf{c})}{D_i(\mathbf{z}_m, \mathbf{z}_{m}^\mathsf{c})}\right)\!, \; \text{subject to}\; \mathbf{z}_m\in \mathcal{Z}_m,
 \end{align} \\[2pt]
where $\mathbf{z}_{m}^\mathsf{c}$ includes all the variable blocks except $\mathbf{z}_m$ i.e., $\mathbf{z}_{m}^\mathsf{c} = \{\mathbf{z}_1,\dots, \mathbf{z}_{m-1},\mathbf{z}_{m+1},\dots, \mathbf{z}_\mathcal{M}\}$.
Now using QT \cite{QT_firstpaper} the $m$th subproblem in equation \eqref{Eq:31} can be expressed in an equivalent form as follows
\begin{align}\label{Eq:32}
    \underset {\mathbf{z}_m,\;\mathbf{y}} {\text{maximize}} \sum\limits_{i=1}^{\mathcal{K}} &f_i\left(2y_i\sqrt{C_i(\mathbf{z}_m, \mathbf{z}_{m}^\mathsf{c})} - y_i^2{D_i(\mathbf{z}_m, \mathbf{z}_{m}^\mathsf{c})}\right),\nonumber \\ 
    \text{subject to}\quad &\mathbf{z}_m\in \mathcal{Z}_m, \quad y_i\in\mathbb{R}, \quad \forall i,
\end{align} \\[1pt]
where $\mathbf{y}$ represents a set of auxiliary variables, denoted as $\{y_1, y_2,\dots,y_\mathcal{K}\}$. 
\end{Proposition} 

We now solve Problem $\textbf{P1A}$ using Proposition~\ref{Proposition-2}. The sum SE in \eqref{SE_bar} can be decomposed as   
\begin{align}\label{Eq:33}
   \overline{\text{SE}}(\boldsymbol{\eta}) = \left(1-\frac{\tau_p}{\tau_c}\right)\sum\limits_{k=1}^{K} \sum\limits_{n=1}^{N}\log_2\left(1+\frac{\Psi_{kn}(\boldsymbol{\eta})}{\Omega_{kn}(\boldsymbol{\eta})}\right),
\end{align}   \\[1pt]
where $\Psi_{kn}(\boldsymbol{\eta})$ and $\Omega_{kn}(\boldsymbol{\eta})$ are the numerator and denominator of the closed-form SINR expression in \eqref{Gamma_cf}. Therefore, Problem $\textbf{P1A}$ can be recast as a sum-of-function-of-ratios optimization problem as
\begin{subequations} \label{P2A}
 \begin{align} \textbf{P2A:}\quad
     \underset {\boldsymbol{\eta}} {\text{maximize}}\quad & \overline{\text{SE}}(\boldsymbol{\eta})\\ 
     \text{subject to} \quad & \eqref{P1A_b},\eqref{P1A_c},\eqref{P1A_d}.
    \end{align}
\end{subequations}
The presence of the fractional ratio objective term in Problem $\textbf{P2A}$ makes it non-convex with respect to the variable $\boldsymbol{\eta}$. We, therefore, divide the set of power control coefficients into $M$ blocks i.e., $\boldsymbol{\eta} = (\boldsymbol{\eta_1}, \dots, \boldsymbol{\eta}_m, \dots, \boldsymbol{\eta}_M)$, where each block comprises the power control coefficients of a specific AP with $\boldsymbol{\eta}_m\in \mathbb{R}^{KN \times 1}$ being the $m$th variable block. Using Proposition $1$, we split Problem $\textbf{P2A}$ into $M$ sub-problems where the $m$th sub-problem can be expressed as 
\begin{subequations} \label{P3A}
 \begin{align} \textbf{P3A:}\quad
     \underset {\boldsymbol{\eta}_{m}} {\text{maximize}}\quad &\overline{\text{SE}}(\boldsymbol{\eta}_{m},\boldsymbol{\eta}_{m}^\mathsf{c}),\\ 
     \text{subject to} \quad &  \eqref{P1A_b},\eqref{P1A_c},\eqref{P1A_d},
\end{align}
\end{subequations} 
where
\begin{align*}\label{Eq:}
    \overline{\text{SE}}(\boldsymbol{\eta}_m, \boldsymbol{\eta}_{m}^\mathsf{c}) = \left(1-\frac{\tau_p}{\tau_c}\right)\sum\limits_{k=1}^{K} \sum\limits_{n=1}^{N}\log_2\left(1+\frac{\Psi_{kn}(\boldsymbol{\eta}_m, \boldsymbol{\eta}_{m}^\mathsf{c})}{\Omega_{kn}(\boldsymbol{\eta}_m, \boldsymbol{\eta}_{m}^\mathsf{c})}\right)\!\!.
\end{align*}
The term $\boldsymbol{\eta}_{m}^\mathsf{c}$ includes the power control coefficients from all other APs except the $m$th AP. To find the optimal solution for Problem $\textbf{P3A}$, it is necessary for $\overline{\text{SE}}(\boldsymbol{\eta}_{m},\boldsymbol{\eta}_{m}^\mathsf{c})$ to be either concave or pseudo-concave. To make $\overline{\text{SE}}(\boldsymbol{\eta}_{m},\boldsymbol{\eta}_{m}^\mathsf{c})$ concave, we apply the concept of fractional programming, where $\overline{\text{SE}}(\boldsymbol{\eta}_{m},\boldsymbol{\eta}_{m}^\mathsf{c})$ is said to be concave only if the ratio terms $\Psi_{kn}(\boldsymbol{\eta}_m, \boldsymbol{\eta}_{m}^\mathsf{c})$ and $\Omega_{kn}(\boldsymbol{\eta}_m, \boldsymbol{\eta}_{m}^\mathsf{c})$ are concave and convex in $\boldsymbol{\eta}_m$ for a fixed $\boldsymbol{\eta}_{m}^\mathsf{c}$ respectively. To decouple the scalar ratios in concave-convex fractional form, we first transform the power control coefficients $\sqrt{{\eta_{mkn}}}$ as ${\xi_{mkn}}$. We next apply QT approach similar to \eqref{Eq:32} to decouple $\overline{\text{SE}}(\boldsymbol{\eta}_{m},\boldsymbol{\eta}_{m}^\mathsf{c})$, and re-cast Problem $\textbf{P3A}$ as follows: 
\begin{subequations} \label{P4A} 
 \begin{align} \textbf{P4A:}\quad \;
     &\underset {\boldsymbol{\xi}_{m},\;\textbf{w}} {\text{maximize}} \quad  {\overline{\text{SE}}^{QT}}(\boldsymbol{\xi}_{m},\boldsymbol{\xi}_{m}^\mathsf{c}), \\ 
     \text{subject to} \; &\sum\limits_{k=1}^{K}\sum\limits_{n=1}^{N} \gamma_{mk}\xi_{mkn}^2\leq 1,\\ 
     & \log_2\left(1+\Gamma_{kn}(\boldsymbol{\xi}_{m})\right) \geq R_{\text{min},kn},\\
     & \sum\limits_{m=1}^{M}\sum\limits_{i=1}^{n-1}\xi_{mki}^2\leq\sum\limits_{m=1}^{M}\xi_{mkn}^2,\;\forall k, \forall n\neq 1,
\end{align}
\end{subequations} 
\begin{align}\label{R_QT}
     &\text{where}\quad{\overline{\text{SE}}^{QT}}(\boldsymbol{\xi}_{m},\boldsymbol{\xi}_{m}^\mathsf{c})= \left(1-\frac{\tau_p}{\tau_c}\right)\sum\limits_{k=1}^{K} \sum\limits_{n=1}^{N}\nonumber \\ &\log_2 \left(1 + 2w_{kn}\sqrt{\Psi_{kn} (\boldsymbol{\xi}_{m},\boldsymbol{\xi}_{m}^\mathsf{c})} - w_{kn}^2{\Omega_{kn} (\boldsymbol{\xi}_{m},\boldsymbol{\xi}_{m}^\mathsf{c})}\right)\!.
\end{align} \vspace{1pt}
The auxiliary variable $w_{kn}\in\mathbb{R}$ is introduced to decouple the SINR term in ${\overline{\text{SE}}(\boldsymbol{\xi}_{m},\boldsymbol{\xi}_{m}^\mathsf{c})}$, and $\mathbf{w}$ represents a set of auxiliary variables $\{w_{11},\dots, w_{KN}\}$. We infer from Problem $\textbf{P4A}$ that, for a given auxiliary variable $w_{kn}$ the function ${\overline{\text{SE}}^{QT}(\boldsymbol{\xi}_{m},\boldsymbol{\xi}_{m}^\mathsf{c})}$ is now concave in $\boldsymbol{\xi}_{m}$ (see \eqref{R_QT}), as the terms $\Psi_{kn} (\boldsymbol{\xi}_{m},\boldsymbol{\xi}_{m}^\mathsf{c})$ and $\Omega_{kn} (\boldsymbol{\xi}_{m},\boldsymbol{\xi}_{m}^\mathsf{c})$ are concave and convex respectively in $\boldsymbol{\xi}_{m}$. For a given $\boldsymbol{\xi} = (\boldsymbol{\xi}_{1}, \dots, \boldsymbol{\xi}_{M})$, the optimal values of the auxiliary variables can be calculated as 
\begin{align}\label{solution}
     \quad w_{kn}^{{*}}\; = \frac{\sqrt{\Psi_{kn}(\boldsymbol{\xi}_{m},\boldsymbol{\xi}_{m}^\mathsf{c})}}{\Omega_{kn}(\boldsymbol{\xi}_{m},\boldsymbol{\xi}_{m}^\mathsf{c})}. 
\end{align} \vspace{1pt}
The power control coefficients for the $m$th AP are optimized iteratively while keeping the powers of other APs fixed. The auxiliary variables are updated until they reach a stationary point. 
The proposed procedure to solve Problem $\textbf{P4A}$ is presented in Algorithm~\ref{Algo:1}.
\begin{algorithm}  
\small
\caption{Successive-QT Based Optimal Power Allocation} \label{Algo:1} 
\begin{algorithmic} [1]
    \State Initialize $\{\boldsymbol{\xi}_{m}\}_{m=1}^M$ with equal power allocation (EPA), and set a maximum number of inner and outer iterations as $J_i$ and $J_o$, where ${\zeta}$ is a pre-defined threshold that decides the error tolerance between two adjacent iterations.   
    
	    \For {$j_o$ = $1,2,\cdots,J_o$}
	        \For {$m$ = $1,2,\cdots,M$}
		        \For {$j_{i}$ = $1,2,\cdots,J_i$} 
			        \State For a given $\boldsymbol{\xi}_m^{j_{i}}$, compute the auxiliary variable $w_{kn}^*$ using \eqref{solution}.
			        \State Solve \textbf{P4A} with respect to the $m$th AP i.e., $\boldsymbol{\xi}_{m}$. 
			        \State Update $\boldsymbol{\xi}_{m}^*$ in the problem.
			        \State Repeat until convergence $\lVert{\boldsymbol{\xi}_{m}^{(j_o,j_{i})} - \boldsymbol{\xi}_{m}^{(j_o,j_{i}-1)}\rVert}^2 \leq {\zeta} $. 
			    \EndFor
		    \EndFor
		    \State Repeat until convergence $\lVert{\boldsymbol{\xi}^{(j_o,j_{i})} - \boldsymbol{\xi}^{(j_o-1,j_{i})}\rVert}^2 \leq {\zeta} $. 
	    \EndFor 
 
    
    \State {Return} $\boldsymbol{\eta}^* = \big(\boldsymbol{\eta}_1^*, \dots, \boldsymbol{\eta}_m^*, \dots, \boldsymbol{\eta}_M^*\big)$, where $\boldsymbol{\eta}_m^*\in \mathbb{R}^{KN \times 1}$ and ${\eta_{mkn}^*}={(\xi_{mkn}^*)}^2,\; \forall m,k,n $.   
\end{algorithmic} 
\end{algorithm} 
\vspace{-5pt}
\subsection{PSO-based RIS Phase Optimization for SE Maximization} 
{We now aim to design the RIS phase matrix $\boldsymbol{\Theta}_r$ for a fixed $\boldsymbol{\eta}$ by solving the sub-problem $\textbf{P1B}$, which is also a non-convex problem. No prior research has tackled the challenges of maximizing SE or EE through the conventional convex optimization approach,  specifically aiming to formulate an optimal RIS phase matrix. {There are some works which considered RIS-aided cell-free MIMO OMA systems \cite{ZhangDZLXZLS21,ZhangD21,42} and have formulated optimization problems to design an optimal RIS phase matrix,  but with different objectives. Reference \cite{ZhangDZLXZLS21} proposed a power minimization approach with an aim to enhance the EE of the system by designing the RIS-based analog beamforming.  The authors in \cite{ZhangD21} formulated a joint precoding design problem for both APs and RISs to maximize the network capacity.  The authors in \cite{42} proposed a non-iterative two-timescale algorithm to maximize the minimum achievable rate (i.e., max-min problem).} {The authors in~\cite{56} summarizes the optimization techniques for RIS-aided wireless communications, including model-based, heuristic,  and machine learning (ML) algorithms. Model-based, heuristic, and ML approaches are compared in terms of stability, robustness, optimality and other parameters,  providing a systematic understanding of these techniques.}

{For the considered SE maximization problem, the objective of the sub-problem $\textbf{P1B}$ contains ratio terms and also the constraints are non-linear, leading to a non-convex fractional programming problem. This makes the problem difficult to solve using traditional convex optimization methods such as successive convex approximation (SCA)\footnote{{Finding the first-order Taylor approximation of each term in the SINR expression with respect to the phase shifts $\mathbf{\Theta}_r$ is quite complex specifically due to the ratio terms containing $\mathbf{\Theta}_r$. To the best of our understanding,  prior literature has not employed SCA technique for designing the RIS phase shifts while addressing such a highly non-convex problem subject to practical conditions including imperfect channels, imperfect SIC and spatial correlation between the RIS elements.}} \cite{53}, semi-definite relaxation, gradient-based approaches, among others.  Exploring unfamiliar environments, particularly in the initial phases where information about the environment is limited,  the reinforcement learning (RL) exploration tends to be essentially random~\cite{RL-3}. This poses a challenge for deploying state-of-the-art RL algorithms in real-world wireless networks.  The lack of performance guarantee during RL exploration means that the network users may have to temporarily suffer from poor Quality of Service (QoS) so that the learning agent can gather information about the deployment for a potentially better RL policy~\cite{RL-3}.  Keeping in mind the high complexity of RIS phase control problems, authors in~\cite{56} highlighted that the heuristic algorithms have lower computational complexity, therefore, they can respond rapidly to real-time network dynamics (Table- XVI, \cite{56}). In addition the heuristic policies can be applied to various scenarios with few requirements on problem formulations.  Our problem is a non-deterministic polynomial (NP)-hard problem due to the tight coupling of RIS phase shifts in each term of the SINR, and heuristic algorithms are particularly useful for solving these problems\cite{55}.}

Particle Swarm Optimization (PSO), a metaheuristic algorithm, is specifically designed to handle such complex optimization problems, which can provide reasonably good solutions in a time frame that is practical for operational use. Computing the first-order derivative of the sum SE with respect to the phase shifts $\boldsymbol{\Theta}_r$ is also quite challenging. Unlike gradient-based methods, PSO does not require computing the derivatives during implementation, allowing faster convergence, especially for non-convex problems with complex and large search spaces \cite{PSO_2}. Therefore, we employ a PSO \cite{PSO_2} based RIS phase shift designing approach to solve Problem $\textbf{P1B}$ in Algorithm~\ref{Algo:2}.

Let $\mathcal{S}$ represent the size of the swarm's population. In an $L$-dimensional $(l=1,2,\dots,L)$ space each particle $i$ $(i=1,2,\dots,\mathcal{S})$ is characterized by its position $\boldsymbol{\theta_i}=(\theta_{i1},\theta_{i2},\dots,\theta_{iL})$ and velocity $\boldsymbol{v_i}=(v_{i1},v_{i2},\dots,v_{iL})$. Here, $\theta_{il}$ takes values in the interval $[0,2\pi)$, and $v_{il}$ is within the range $[v_{\text{min}},v_{\text{max}}]$. Define the objective function as $\overline{\text{SE}}_{fit}(\boldsymbol{\Theta}_r) = -\left(1-\frac{\tau_p}{\tau_c}\right) \sum_{k=1}^{K}\sum_{n=1}^{N}\log_2 \left(1 +\Gamma_{kn}(\boldsymbol{\eta,\Theta_r})\right)$, and calculate the fitness value of each particle as $\overline{\text{SE}}_{fit}\boldsymbol{(\theta_i)}$. Represent the personal best position of each particle $i$ as $\boldsymbol{\theta_{\text{best},i}}$, and the global best position of the swarm as $g_{\text{best}}$, where $\boldsymbol{\theta_{\text{best},i}}=(\theta_{\text{best},i1},\theta_{\text{best},i2},\dots,\theta_{\text{best},iL})$ and $g_{\text{best}}=(g_{\text{best},1},g_{\text{best},2},\dots,g_{\text{best},L})$. Initially, $\boldsymbol{\theta_{\text{best},i}}$ is set as $\boldsymbol{\theta_i}$, and $g_{\text{best}}$ is determined as the position with the smallest fitness value among all particles, i.e., $\min\{{\overline{\text{SE}}_{fit}\boldsymbol{(\theta_1)},\overline{\text{SE}}_{fit}\boldsymbol{(\theta_2)},\dots,\overline{\text{SE}}_{fit}\boldsymbol{(\theta_\mathcal{S})}}\}$. Update the velocity $v_i$ of each particle using \eqref{velocity-update}, and the position $\boldsymbol{\theta_i}$ using \eqref{position-update}. The constants $c_1$ and $c_2$ in \eqref{velocity-update} are the acceleration constants that impact the maximum step size and are typically chosen within the range $(0\leq c_1,c_2\leq 2)$ for more accurate results. $u_1$ and $u_2$ are randomly generated within the range $[0,1]$. A constriction factor $\kappa$ is introduced in \eqref{velocity-update} to achieve faster convergence, which is expressed as $\kappa=2/|2-\omega-\sqrt{\omega^2-4\omega}|$, where $\omega=c_1+c_2$ and $\omega>4$ \cite{PSO_2}.\\

\begin{algorithm}  
\small
\caption{Optimal RIS Phase Designing Using PSO}\label{Algo:2}
\begin{algorithmic}[1]
\State Initialize: $\mathcal{S}, T, \kappa, u_{1}, u_{2},c_{1}, c_{2},t=1$.
    \State Initialize the position $\boldsymbol{\theta_{i}}^{(0)}$ and velocity $\mathbf{v_{i}}^{(0)}$ for the $i$th particle in the population. Assume $\boldsymbol{\theta}_{best_{i}}^{(0)}=\boldsymbol{\theta}_{i}^{(0)}$ and determine $\overline{\text{SE}}_{fit}(\boldsymbol{\theta}_{best_{i}}^{(0)})$, $\forall i=1,2,\cdots,\mathcal{S}$.
    \State Compute the initial best position of the swarm, denoted as $\boldsymbol{\theta}_{best}^{(0)}$, by evaluating  $\overline{\text{SE}}_{fit}(\boldsymbol{\theta}_{best}^{(0)})=\text{min}\left\{\overline{\text{SE}}_{fit}(\boldsymbol{\theta}_{best_{1}}^{(0)}),\cdots\!,\overline{\text{SE}}_{fit}(\boldsymbol{\theta}_{best_{\mathcal{S}}}^{(0)})\right\}$. Assign $\mathbf{g}_{best}^{(0)}=\boldsymbol{\theta}_{best}^{(0)}$.
    \While {$t<$ $T$}
    
       \For{$i = 1,2,\cdots,\mathcal{S}$}
            \State Use the following equations to update the velocity and position vectors: 
            \begin{flalign}\label{velocity-update}
            \mathbf{v}_{i}^{(t+1)}
            =\kappa(\mathbf{v}_{i}^{(t)}+c_{1}u_{1}(\boldsymbol{\theta}_{best_{i}}^{(t)}-\boldsymbol{\theta}_{i}^{(t)})
              +  c_{2}u_{2}(\mathbf{g}_{best}^{(t)}-\boldsymbol{\theta}_{i}^{(t)})), 
            \end{flalign}
            \begin{flalign}\label{position-update}
            \boldsymbol{\theta}_{i}^{(t+1)}=\boldsymbol{\theta}_{i}^{(t)}+\mathbf{v}_{i}^{(t+1)}. 
            \end{flalign}
            \State For the updated position of each particle, calculate the fitness value. 
            \State For each particle $i$, update the personal best 
            \If {$\overline{\text{SE}}(\boldsymbol{\theta}_{i}^{(t+1)})<\overline{\text{SE}}(\boldsymbol{\theta}_{best_{i}}^{(t)})$}
            \State $\boldsymbol{\theta}_{best_{i}}^{(t+1)}=\boldsymbol{\theta}_{i}^{(t+1)}$,
            \Else
            \State $\boldsymbol{\theta}_{best_{i}}^{(t+1)}=\boldsymbol{\theta}_{best_{i}}^{(t)}$.
            \EndIf
        \EndFor
        \State For the current instance, find the best position of the swarm:  
        \begin{equation}
        \overline{\text{SE}}(\boldsymbol{\theta}_{best}^{(t+1)})=\min\left\{\overline{\text{SE}}(\boldsymbol{\theta}_{best_{1}}^{(t+1)}),\cdots,\overline{\text{SE}}(\boldsymbol{\theta}_{best_{L}}^{(t+1)})\right\}\nonumber.
        \end{equation}
        \State Find the swarm's optimum position $\mathbf{g}_{best}^{(t+1)}$
        \If {$\overline{\text{SE}}(\boldsymbol{\theta}_{best}^{(t+1)})<\overline{\text{SE}}(\mathbf{g}_{best}^{(t)})$}
        \State $\mathbf{g}_{best}^{(t+1)}=\boldsymbol{\theta}_{best}^{(t+1)}$,
        \Else
        \State $\mathbf{g}_{best}^{(t+1)}=\mathbf{g}_{best}^{(t)}$.
        \EndIf
        \State $t \leftarrow t+1$.
    \EndWhile
    \State $\mathbf{g}_{best}^{(T)}$ is the optimized phase vector.
\end{algorithmic}
\end{algorithm} 

We now summarize the joint optimization framework to maximize the sum SE in Algorithm~\ref{Algo:3}.

\vspace{-10pt}
\subsection{Computational Complexity}  
\vspace{-2pt}
{In the successive-QT algorithm, each innermost iteration $j_2$ performs the calculation of auxiliary variables $\mathbf{w}$ and solves Problem $\textbf{P4A}$. The computation of $\mathbf{w}$ in Step $5$ is straightforward and does not add much complexity. The overall computational complexity of Algorithm \ref{Algo:1} is mainly determined by the complexity of solving Problem $\textbf{P4A}$ in each iteration, which involves a total $KN$ number of real optimization variables and $(KN+2)$ linear constraints. Therefore, the computational complexity at each iteration can be written as $\mathcal{O}((2KN+2)^{3/2}(KN)^{2})$ \cite{Complexity-analysis}. 
Hence for all the $M$ APs, the overall complexity of the proposed approach is $\mathcal{O}(M((2KN+2)^{3/2}(KN)^{2}))$.}   

{The computational complexity of the PSO algorithm depends on the calculations required to update the velocity \eqref{velocity-update}  and position \eqref{position-update}  of particles, as well as the number of fitness function evaluations. In each iteration of the algorithm, $L\mathcal{S}$ number of multiplications are required to update the velocity vector, where $L$ is the number of RIS elements and $\mathcal{S}=\min\{100,10L\}$ represents the population size. Additionally, the fitness function is evaluated $\mathcal{S}$ times in each iteration. Therefore, the overall complexity of the PSO algorithm can be expressed as $\mathcal{O}(L\mathcal{S}T+T\mathcal{S})$, where $T$ denotes the number of PSO iterations.} 
\begin{algorithm}  
\small
\caption{Joint Power and Phase Optimization Framework} \label{Algo:3} 
\begin{algorithmic} [1]
    \State Divide the joint optimization Problem \textbf{P1} into two subproblems i.e., \textbf{P1A} and \textbf{P1B}.    
    \State Initialize the power control coefficients $\boldsymbol{\eta}$ with equal power allocation (EPA) and RIS phase matrix $\boldsymbol{\Theta}_r$ with randomly generated phase shifts. 
    \State Solve subproblem \textbf{P1A} to optimize $\boldsymbol{\eta}$ for a fixed $\boldsymbol{\Theta}_r$ using successive-QT in Algorithm~ \ref{Algo:1}.
    \State Based on the optimized $\boldsymbol{\eta}$, solve subproblem \textbf{P1B} to optimize $\boldsymbol{\Theta}_r$ using PSO in Algorithm~ \ref{Algo:2}.
    \State Let, $\overline{\text{SE}}$ denotes the objective function of the formulated problem in \textbf{P1}. The algorithm converges {if}, in the $i$th iteration, $\overline{\text{SE}}^{(i)} - \overline{\text{SE}}^{(i-1)}\leq \varepsilon$; {else, goto Step $3$ and repeat.} Here, $\varepsilon$ is the predefined threshold between two consecutive iterations.
\end{algorithmic}
\end{algorithm}  
\vspace{-18pt}
\subsection{Convergence Analysis}
\vspace{-2pt}
{We now analyze the convergence of the proposed algorithm using the following lemma.
\begin{lemma} \vspace{-5pt} 
Consider the optimization problems \(\mathbf{P1A}\), \(\mathbf{P2A}\), \(\mathbf{P3A}\), and \(\mathbf{P4A}\) with corresponding objective functions \(f_{1A}(\boldsymbol{\eta})\), \(f_{2A}(\boldsymbol{\eta})\), \(f_{3A}(\boldsymbol{\eta}_m, \boldsymbol{\eta}_{m}^\mathsf{c})\), and \(f_{4A}(\boldsymbol{\xi}_m, \boldsymbol{\xi}_{m}^\mathsf{c}, \mathbf{w})\). Let \(j_i^m\) and \(j_o\) denote the inner and outer loop iterations, respectively, for the \(m\)th AP. Then,
\begin{enumerate} \addtolength{\leftskip}{-5pt}
\item At each iteration \((j_o, j_i^m)\), the optimization variables \(\bar{\boldsymbol{\xi}}_m^{\;(j_o,\; j_i^m)}\) and auxiliary variable \(\mathbf{w}^{(j_o,\; j_i^m)}\) satisfy the following sequence of equalities:
\begin{align*}
   f_{1A}\big(\boldsymbol{\eta}^{(j_o+1,\; j_i^m)}\big) &= f_{2A}\big(\boldsymbol{\eta}^{(j_o+1,\; j_i^m)}\big) = f_{3A}\big(\bar{\boldsymbol{\eta}}_m^{\;(j_o+1,\; j_i^m)}\big) \nonumber \\ &= f_{4A}\big(\bar{\boldsymbol{\xi}}_m^{\;(j_o+1,\; j_i^m)}, \mathbf{w}^{(j_o+1,\; j_i^m)}\big).
\end{align*}
\item The objective function \(f_{4A}\big(\bar{\boldsymbol{\xi}}_m^{\;(j_o+1,\; j_i^m)}, \mathbf{w}^{(j_o+1,\; j_i^m)}\big)\) is monotonically non-decreasing with respect to both the inner and outer loop iterations. 
\item The overall objective function \(f_{1A}\big(\boldsymbol{\eta}^{(j_o,\; j_i^m)}\big)\) converges to a stationary point \(\boldsymbol{\eta}^*\) as the number of iterations increases.
\item For the original problem \(\mathbf{P1}\) with objective function \(f_{\text{SE}\;}\big(\boldsymbol{\eta}, \boldsymbol{\Theta}_r\big)\), the sequence \(f_{\text{SE}\;}\big(\boldsymbol{\eta}^{(t)}, \boldsymbol{\Theta}_r^{(t)}\big)\) is monotonically non-decreasing and converges to a sub-optimal point \((\boldsymbol{\eta}^*, \boldsymbol{\Theta}_r^*)\).
\end{enumerate}
\end{lemma}\vspace{-3pt}
\begin{proof}
The proof directly follows from the application of the equivalent objective conditions \cite{QT_firstpaper}, equivalent solution properties, first-order optimality conditions \cite{Convergence_2}, and convergence characteristics of the underlying algorithms.   
\end{proof} 
}

\vspace{-5pt}
\begin{Remark}
It is important to note that most of the existing work~\cite{abs-2104-08648,42} on RIS-aided cell-free systems are specific instances of the general scenario studied in this work. Therefore, our proposed joint optimization framework can serve as a comprehensive solution to maximize the SE in most of these existing works. 
\end{Remark} 

\begin{figure*}[htbp]
	\centering	
        \begin{subfigure}[b]{0.30\linewidth}              \hspace{-30pt}\includegraphics[width=1.2\linewidth,height=1\linewidth]{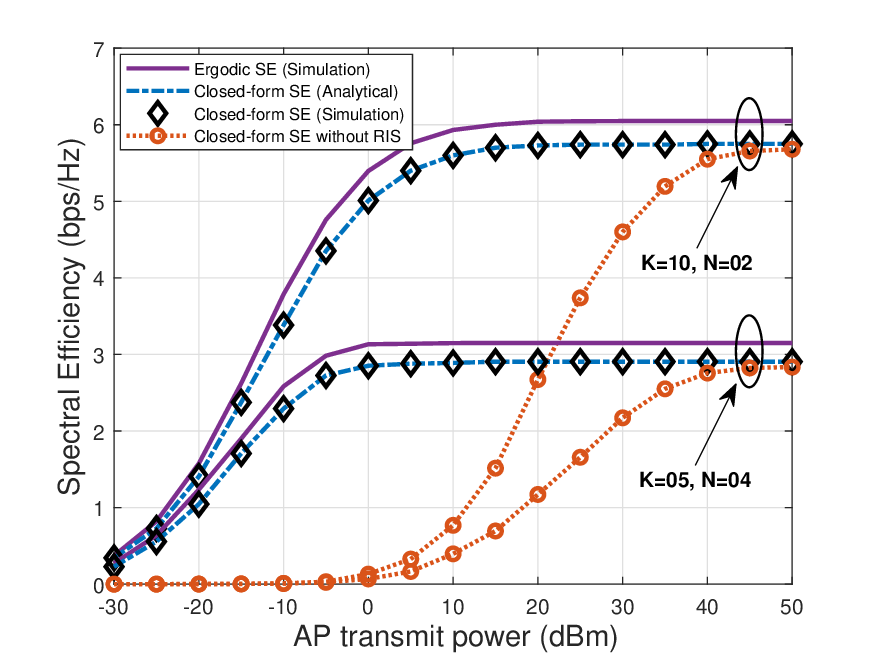}\vspace{-5pt}
                \caption{\small}
                \label{fig:se_pmax}
        \end{subfigure}\vspace{-3pt}
        \begin{subfigure}[b]{0.30\linewidth}	   \hspace{-15pt}\includegraphics[width=1.2\linewidth,height=1\linewidth]{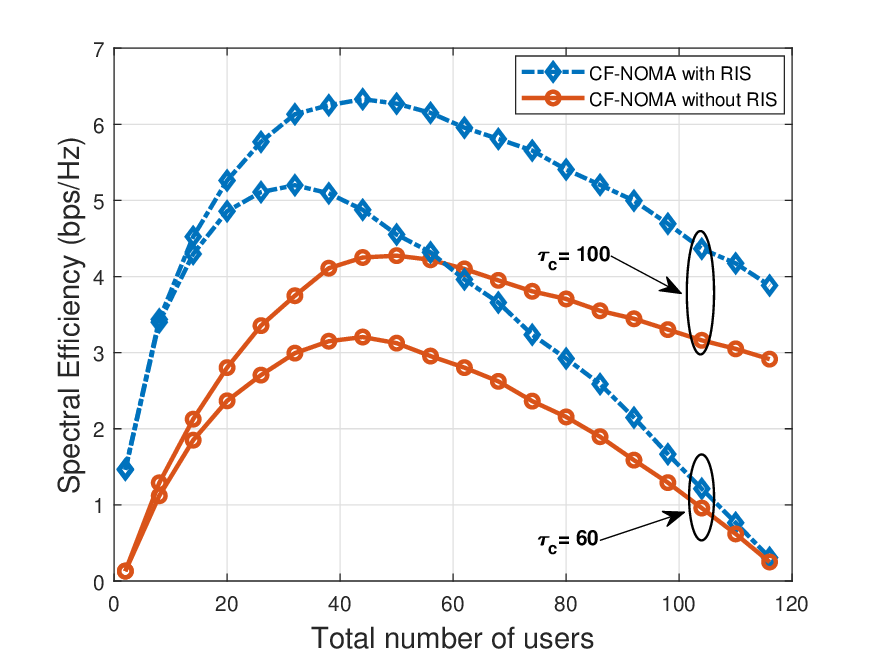}\vspace{-5pt}
              \caption{\small}
	        \label{fig:se_users}
	\end{subfigure}\vspace{-3pt} 
        \begin{subfigure}[b]{0.30\linewidth}	   \hspace{0pt}\includegraphics[width=1.2\linewidth,height=1\linewidth]{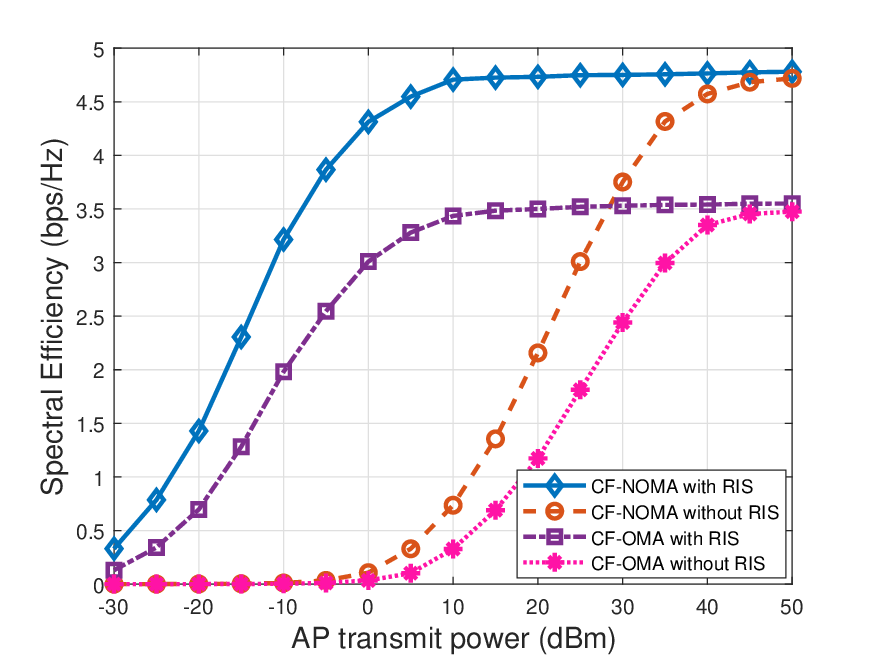}\vspace{-5pt}
              \caption{\small}
	       \label{fig:se_pmax_NOMA_OMA}
	\end{subfigure} \vspace{-3pt} 
	\caption{\small {a) Spectral efficiency versus AP transmit power, b) Spectral efficiency versus the total number of users, c) Spectral efficiency comparison of cell-free NOMA versus cell-free OMA. \\[-30pt]}}
\end{figure*} 
\vspace{-22pt}
\section{Simulation Results} 
In this section, we evaluate the performance of the RIS-assisted cell-free mMIMO NOMA system using extensive simulations. {For simulation, we consider a large geographical area of $(1000 \times 1000)$ m$^2$, where the APs and users are uniformly distributed. The four vertices of the considered region in terms of $(x,y)$ coordinates are $[-500,500]$ m, $[-500,-500]$ m, $[500,-500]$ m and $[500,500]$ m. The RIS is located at $(x,y)= (0,100)$~m in the considered simulation area. The users are deployed outside a semi-circle of radius $d_\text{min}=100$~m from the RIS.  We consider the large-scale path-loss parameters according to the three-slope propagation model as $\beta_{mk}=\text{PL}_{mk} + \sigma_{sh} z_{mk}$ (\text{dB})~\cite{NgoAYLM17}, where $\text{PL}_{mk}$ is the distance dependent path-loss which depends on the distance between the transmitter and the receiver, $\sigma_{sh}z_{mk}$ represents the shadow fading with standard deviation $\sigma_{sh}$, and $z_{mk}\sim \mathcal {CN}(0,1)$. The shadow fading has a log-normal distribution with a standard deviation ($\sigma_{sh}$) equal to $14$ dB and $4$ dB for the direct and indirect links,  respectively. The distance thresholds for the three slopes are $10$ m and $50$ m. The height of the APs, RIS, and users is $15$ m, $30$ m, and $1.65$ m, respectively~\cite{abs-2104-08648}.} 

{We consider the spatial correlation model from~\cite{DBLP:IRS_emil_sir} to generate the covariance matrices. For clustering, we follow the mechanism where the users with the smallest distance from each other are paired \cite{BasharCBNHX20}.  We continue pairing the nearest users until all the users are group into clusters.\footnote{{Future work can consider designing either non machine learning or machine learning based clustering algorithms to further improve the SE \cite{Clustering}.}} For OMA baseline, we accommodate the users via SDMA \cite{abs-2104-08648}. 
The power control coefficient of the $n$th user in the $k$th cluster for the $m$th AP,  is taken as ${\eta_{mkn}} = \big(\sum_{k'=1}^{K} \gamma_{mk'} \big)^{-1}$,  which is obtained from \eqref{power_constraint} to satisfy the total transmit power constraint~\cite{NgoAYLM17}. 
The other system parameters are considered as follows \cite{NgoAYLM17,abs-2104-08648}: i) carrier frequency $f_c= 1.9$~GHz, ii) wavelength $\lambda_c = 15.8$~cm, iii) bandwidth $B=20$~MHz, iv) horizontal and vertical width of each RIS element $d_H=d_V=\lambda_c/4$ and $d_V=\lambda_c/4$, v) noise variance~$=-92$~dBm, corresponding to a noise figure of $9$~dB, vi) pilot power~$=20$~dBm, and vii) $R_{\text{min},kn}=0.1$~bps/Hz, $\forall k, n$.  
In the PSO algorithm, the following parameters are considered \cite{PSO_2}: i) $\mathcal{S} = \min\{100, 10L\}$, ii) $v_{\text{max}} = 2\pi$, iii) $v_{\text{min}} = 0$, iv) $c_1 = 2.05$, v) $c_2 = 2.05$, vi) $\kappa = 0.7298$, and vii) $T = 5L$. }
 
\vspace{-15pt}
\subsection{Validation of Closed-Form Spectral Efficiency Expression Derived in Theorem-\ref{Theorem 1}} 
Fig. \ref{fig:se_pmax} illustrates the SE performance for different transmit powers $\rho_d$ while serving a total $20$ number of users with two different setups, i.e.,  $K=10$ clusters with $N=2$ users per cluster and $K=5$ clusters with $N=4$ users per cluster. We set $M= 64$ APs,  $L= 64$ RIS elements, and the coherence interval, $\tau_c=100$. {We observe from Fig.~\ref{fig:se_pmax} that the derived closed-form SE expression using~\eqref{Gamma_cf} (labeled as “Closed-form SE (Analytical)”) exactly matches with the SE obtained using~\eqref{Gamma} (labeled as “Closed-form SE (Simulation)”), which confirms the accuracy of the derivation. We also observe that the derived SE expression using~\eqref{Gamma_cf} in closely follows the ergodic SE in~\eqref{ergodicsumrate} (labeled as ``Ergodic SE (Simulation)"), which validates Theorem~\ref{Theorem 1}.} {To show the advantages of using RIS, we also plot the SE in Fig.~\ref{fig:se_pmax} in the absence of RIS (labeled as ``Closed-form SE without RIS"). We clearly observe that for transmitting power $\rho_d< 0$~dBm, the SE without RIS is almost zero due to the high path loss and shadowing. On the contrary, with the integration of RIS, we get a significant boost in the SE value even with very low transmit power.  Therefore, this justifies using the RIS when there is a weak direct link between the user and the AP.  We also observe that for higher transmit power, i.e., $\rho_d\geq 40$~dBm, the SE obtained with and without RIS are equal. This is because at higher transmit powers, the direct link becomes dominant over the RIS-aided link. As we increase $N$, the SINR at each user decreases due to the increase in intra-cluster interference, which in turn reduces the SE.} 

\begin{figure*}[htbp]
	\centering	
        \begin{subfigure}[b]{0.30\linewidth}              \hspace{-30pt}\includegraphics[width=1.2\linewidth,height=1\linewidth]{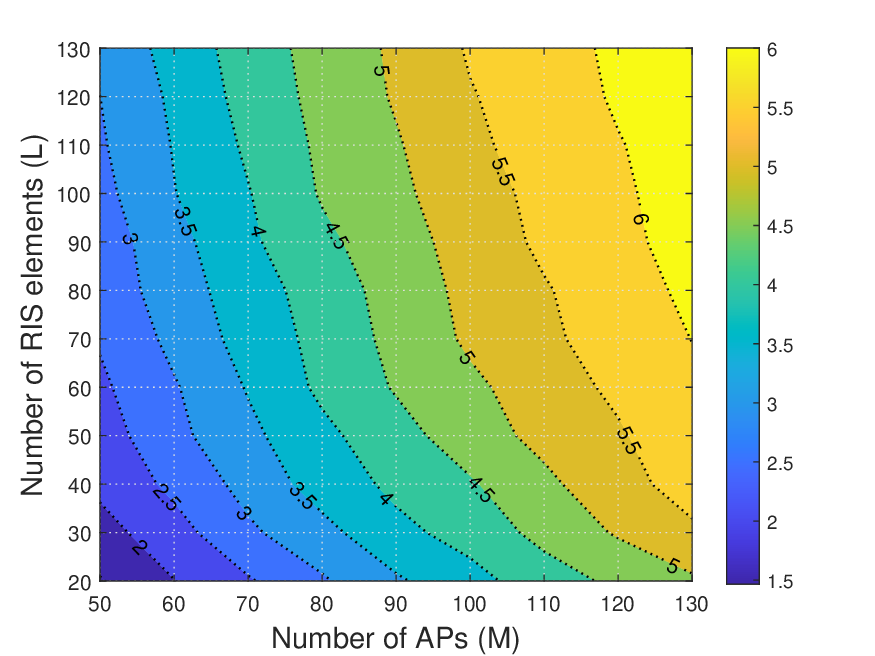}\vspace{-5pt}
                \caption{\small}
                \label{fig:se_ap_ris_contour}
        \end{subfigure}\vspace{-3pt}
        \begin{subfigure}[b]{0.30\linewidth}	   \hspace{-15pt}\includegraphics[width=1.2\linewidth,height=1\linewidth]{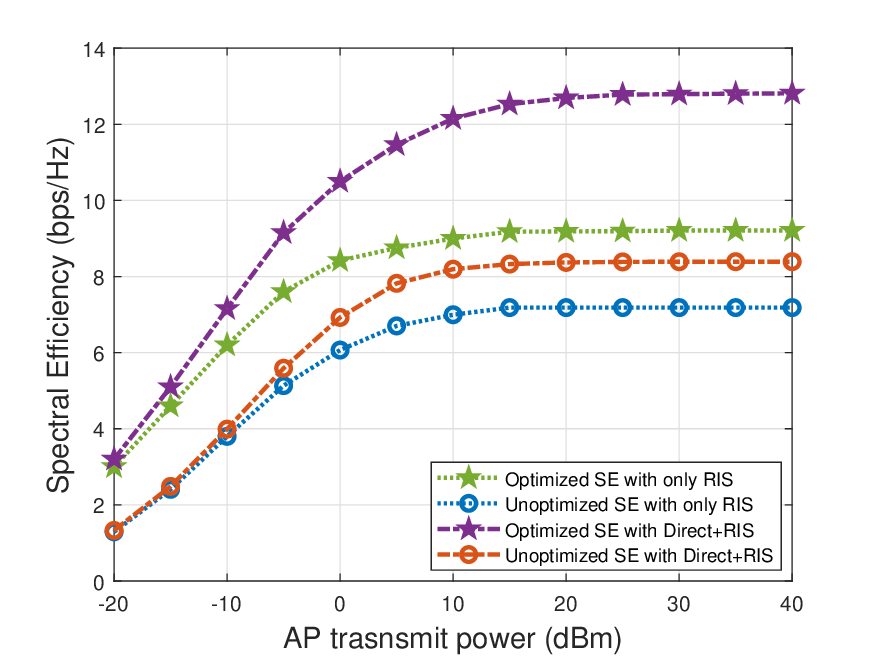}\vspace{-5pt}
              \caption{\small}
	        \label{fig:se_power_opt_1}
	\end{subfigure}\vspace{-3pt} 
        \begin{subfigure}[b]{0.30\linewidth}	   \hspace{0pt}\includegraphics[width=1.2\linewidth,height=1\linewidth]{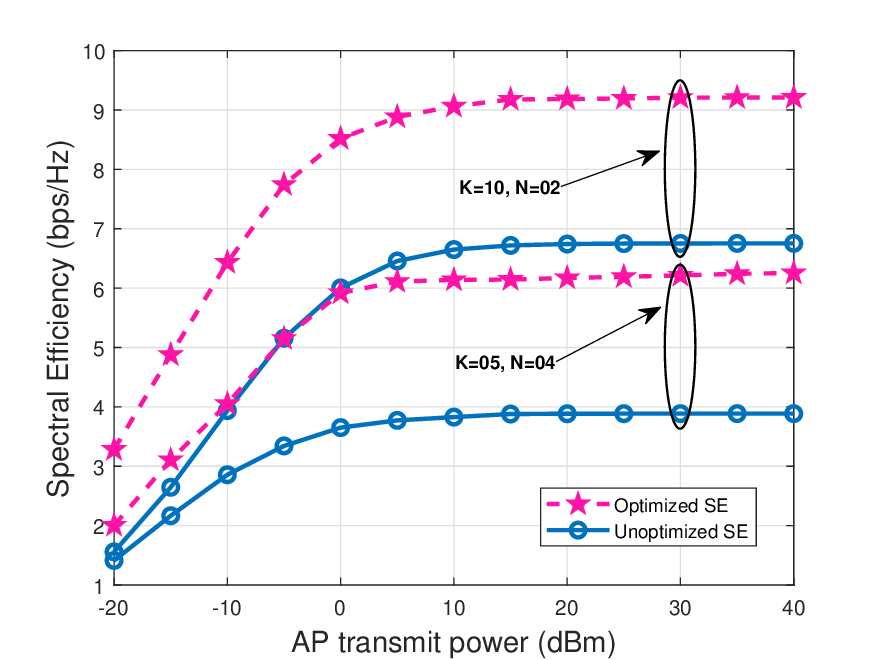}\vspace{-5pt}
              \caption{\small}
	       \label{fig:se_power_opt_2}
	\end{subfigure} \vspace{-3pt} 
	\caption{\small a) Impact of APs and RIS elements on the spectral efficiency, b) Spectral efficiency versus AP transmit power with the proposed algorithm, c) Spectral efficiency versus AP transmit power with optimization for different scenarios.\\[-30pt]}
\end{figure*}  
\vspace{-10pt}
\subsection{Impact of the Number of Users on the Achievable SE} 
Fig.~\ref{fig:se_users} demonstrates the SE versus total number of users ($KN$) for two different coherence intervals i.e., $\tau_c= 100$ and $\tau_c= 60$. We set $M= 64$ APs, $L= 64$ RIS elements, $\rho_d= 20$ dBm and $N= 2$ users in each cluster. We observe that with the increase in the number of users, the SE value initially increases and then decreases after a certain number of users due to an increase in channel estimation overhead. 
{We note that for a coherence interval of $\tau_c= 60$ and $\tau_c= 100$, the maximum SE achieved by the users with RIS-assisted links increases by $62\%$ and $48\%$, respectively, as compared to the scenario when the RIS-assisted links are absent. 
We see that the advantage of using the RIS starts to diminish as we keep increasing the number of users.  
We also observe that as the number of users increases, the inter-cluster interference becomes more prominent, leading to a decrease in the overall SE of the system.} In addition, with high coherence interval $\tau_c$, the system is capable of serving more users with better SE.     
\vspace{-5pt}
\subsection{{Comparison of Cell-Free NOMA and Cell-Free OMA}}
Fig.~\ref{fig:se_pmax_NOMA_OMA} compares the SE of RIS-assisted cell-free NOMA and cell-free OMA systems by varying the transmit power of each AP. We choose $M= 64$ APs, $L= 128$ RIS elements, $K= 20$ clusters with $N= 2$ users per cluster and a coherence interval of $\tau_c= 50$. The results show that NOMA outperforms OMA in terms of SE for the given parameters for both the direct and RIS-assisted links. {We also observe that in the lower transmit power region, i.e., for $\rho_d\leq 0$ dBm, the SE is almost zero when the RIS-assisted links are absent in both NOMA and OMA systems. More specifically, for $\rho_d\geq40$ dBm, there is a relative improvement of $36\%$ in the SE with NOMA when compared to its OMA counterpart.}
 
\vspace{-10pt}
\subsection{Joint Impact of Number of APs and RIS Elements on SE} 
Fig.~ \ref{fig:se_ap_ris_contour} investigates the impact of the number of APs $M$ and RIS elements $L$ on the downlink SE of the RIS-assisted cell-free NOMA system. We set $\tau_c= 100$, $K= 10$ clusters with $N= 2$ users per cluster and $\rho_d= -10$~dBm. Each point in the contour plot represents a fixed SE value and reveals its dependence on $M$ and $L$  for achieving the desired SE. We note that for any fixed $M$, the SE increases roughly around $2$ bps/Hz when $L$ is increased from $20$ to $130$. Similarly, for any fixed $L$, the SE increases around $3$ bps/Hz when $M$ is increased from $50$ to $130$.  We observe that for the lower number of APs, the impact of RIS elements on the SE is more significant.  More specifically,  there is a relative improvement of $75\%$ and $36\%$ in the SE for $M=60$ and $M=100$, respectively, when we increase the number of RIS elements from $20$ to $100$. 

\vspace{-5pt}
\subsection{Effect of Joint Power and Phase Optimization on the SE} 
Fig. \ref{fig:se_power_opt_1} illustrates the SE versus AP transmit power to numerically investigate the efficacy of the proposed solution to maximize the sum SE of the system with joint power allocation and phase shifts optimization using Algorithm~\ref{Algo:3}. We set $M = 64$ APs, $L = 64$ RIS elements, $K = 10$ clusters, and $N = 2$ users in each cluster. We observe that the SE with the proposed algorithm outperforms the SE without optimization for both scenarios, i.e., with only RIS and direct plus RIS-assisted links. For $\rho_d\geq10$ dBm, the \textcolor{black}{proposed algorithm} results in $56\%$ better SE than its EPA counterpart. For the scenario when the direct link is absent, the \textcolor{black}{proposed algorithm} results in $30\%$ higher SE than EPA. We also observe that the SE obtained with \textcolor{black}{optimization} in the absence of direct link \textcolor{black}{outperforms} the SE obtained with direct link for~EPA.     

Fig. \ref{fig:se_power_opt_2} depicts the impact of intra-cluster interference on the SE optimization in the presence of imperfect SIC. We set $M = 64$ APs, $L = 64$ RIS elements, and have a total of $20$ users with two different scenarios: 1) $K=10$ clusters with $N=2$ users per cluster and 2) $K=5$ clusters with $N=4$ users per cluster. To highlight the benefits of utilizing RIS, we assume that direct links between users and APs are not available. We see that despite significant intra-cluster interference in {Scenario~2}, the SE achieved with the proposed algorithm demonstrates superior performance and produces nearly the same SE obtained in Scenario~1 without optimization. Moreover, for $\rho_d\geq 10$ dBm, we can observe that there is a noticeable gap of almost $3$ bps/Hz between the optimized and unoptimized SE in both scenarios.  

\vspace{-10pt}
\subsection{NOMA-OMA Comparison with the Proposed Algorithm}
Fig. \ref{fig:se_opt_NOMA_OMA} compares the sum SE of the cell-free NOMA and cell-free OMA systems with and without optimization. We choose $M= 64$ APs, $L= 64$ RIS elements, $K= 20$ clusters, $N= 2$ users per cluster, and a coherence interval of $\tau_c= 50$. The results show that the NOMA outperforms OMA in terms of SE for the given parameters in both scenarios, with and without optimization. 
In comparison to equal power allocation and random phase shifts, when the power allocation and phase shifts are optimized, the SE gap between NOMA and OMA notably improves.  With the proposed algorithm, we observe that the SE for NOMA is almost doubled for $\rho_d\geq10$ dBm, whereas for OMA, there is  an improvement of around~$60\%$.  
\begin{figure*}[htbp]
	\centering	
        \begin{subfigure}[b]{0.30\linewidth}	   \hspace{-30pt}\includegraphics[width=1.2\linewidth,height=1\linewidth]{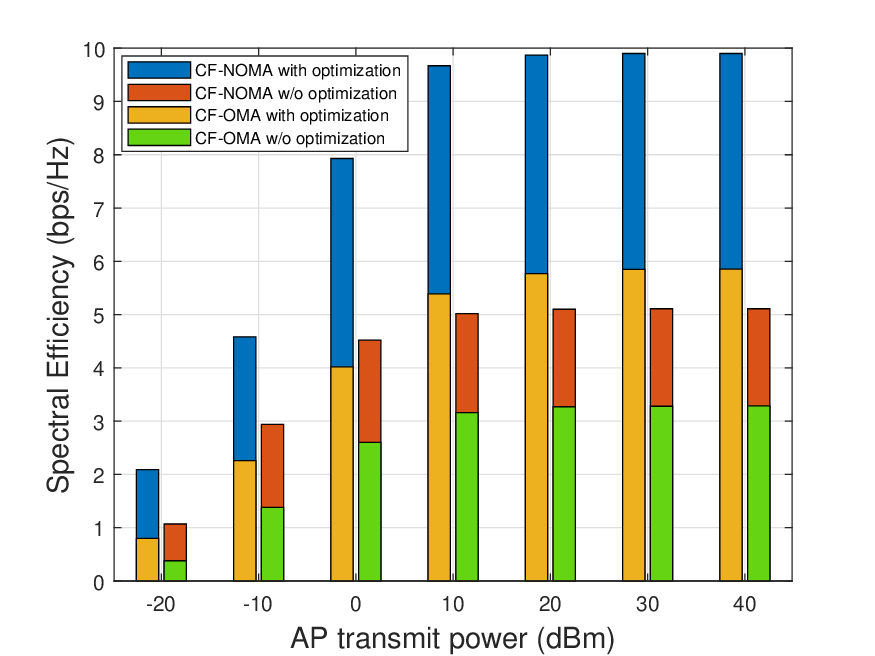}\vspace{-5pt}
              \caption{\small}
	        \label{fig:se_opt_NOMA_OMA}
	\end{subfigure}\vspace{-8pt}
        \begin{subfigure}[b]{0.30\linewidth}	   \hspace{0pt}\includegraphics[width=1.2\linewidth,height=1\linewidth]{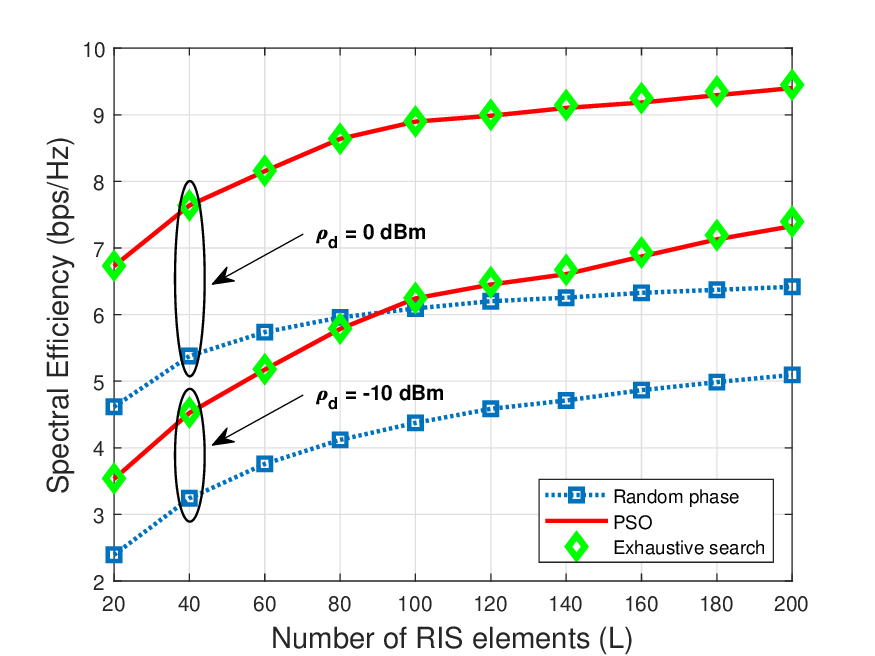}\vspace{-5pt}
              \caption{\small}
	       \label{fig:8}
	\end{subfigure} \vspace{-2pt} 
	\caption{\small a) Cell-free NOMA versus cell-free OMA with and without optimization, b) {Spectral efficiency versus RIS elements with random and optimal phase~shifts.\\[-20pt]}}
\end{figure*} 

\begin{figure*}[htbp]
	\centering	
        \begin{subfigure}[b]{0.30\linewidth}              \hspace{-30pt}\includegraphics[width=1.2\linewidth,height=1\linewidth]{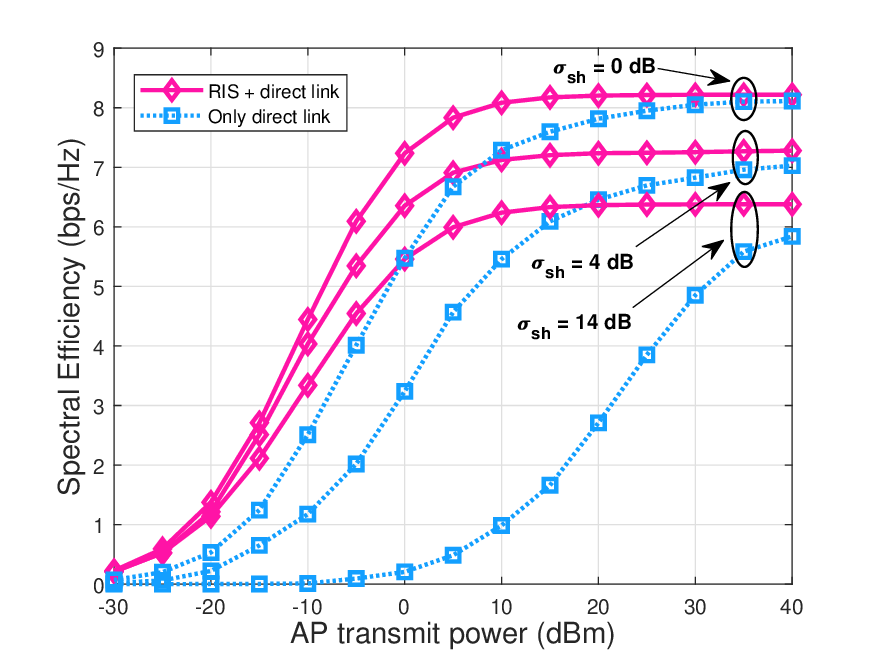}\vspace{-5pt}
                \caption{\small}
                \label{fig:10}
        \end{subfigure}\vspace{-3pt}
        \begin{subfigure}[b]{0.30\linewidth}	   \hspace{-15pt}\includegraphics[width=1.2\linewidth,height=1\linewidth]{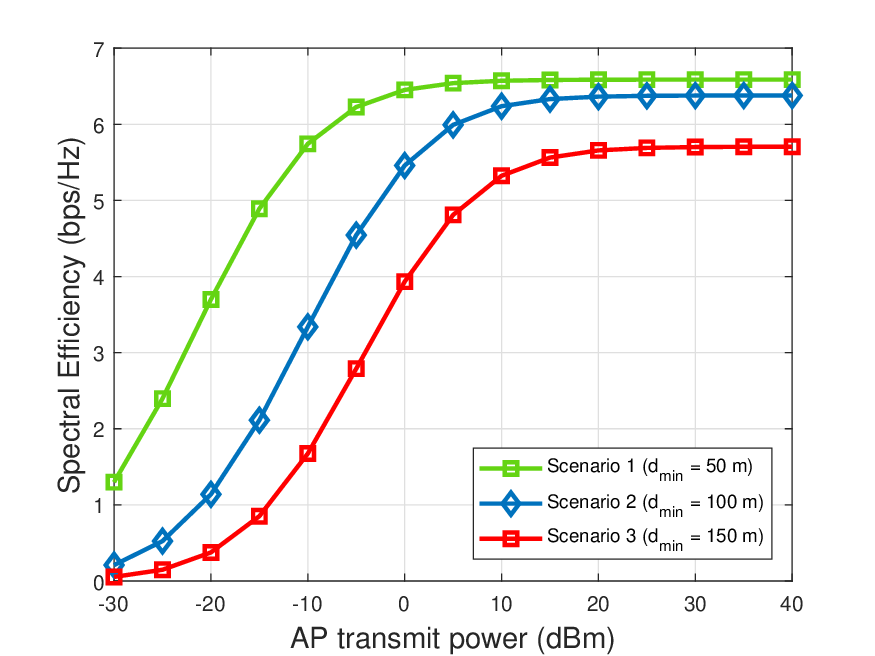}\vspace{-5pt}
              \caption{\small}
	        \label{fig:11}
	\end{subfigure}\vspace{-3pt} 
        \begin{subfigure}[b]{0.30\linewidth}	   \hspace{0pt}\includegraphics[width=1.2\linewidth,height=1\linewidth]{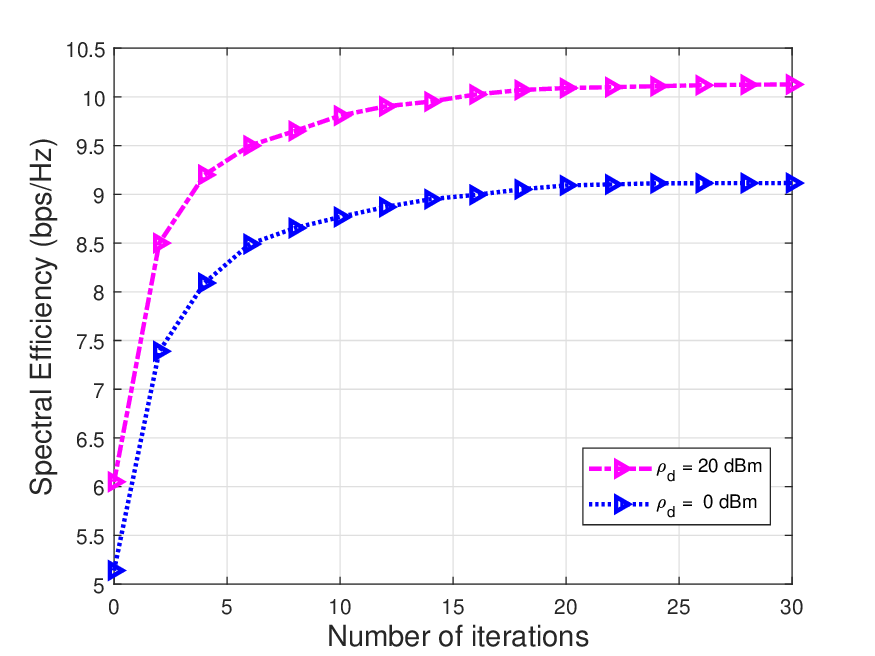}\vspace{-5pt}
              \caption{\small}
	       \label{fig:12}
	\end{subfigure} \vspace{-3pt} 
	\caption{\small {a) Spectral efficiency versus AP transmit power for different direct path scenarios, b) Spectral efficiency versus AP transmit power for different user distribution scenarios, {c) SE versus number of iterations for Algorithm $3$}.\\[-30pt]}}
\end{figure*}  

\vspace{-10pt}
\subsection{{Impact of the Number of RIS Elements on the SE with Phase Optimization}}
Fig. \ref{fig:8} demonstrates the impact of RIS elements on the SE with optimal phase shifts for two different transmit power levels, i.e., $\rho_d = -10$ dBm and $\rho_d = 0$ dBm. We set $M = 64$ APs, $K = 10$ clusters, and $N = 2$ users in each cluster and a coherence interval of $\tau_c=100$. {To evaluate the effectiveness of the proposed PSO algorithm, we plot in Fig. \ref{fig:8} the SE obtained using: i) PSO algorithm (labeled as “PSO”), ii) random phase allocation (labeled as “Random phase”), and iii) exhaustive search (labeled as “Exhaustive search”).  We, similar to \cite{ES1,ES4},  for benchmark purpose took discrete values of the phase co-efficients and then performed exhaustive search by generating 100000 number of random phase vectors. Finally, we choose the phase vector as the one with the maximum SE. We observe the proposed PSO algorithm and exhaustive search exhibit close proximity, indicating that the proposed PSO algorithm effectively converges. We also see that the SE increases with the number of RIS elements $L$ for both random and optimized phase shifts due to an array gain provided by a larger number of RIS elements.  Additionally, for $L\geq 160$, we observe there is a notable gap of almost $2$ bps/Hz and $3$ bps/Hz between the SE achieved with PSO and random phase shifts for $\rho_d = -10$ dBm and $\rho_d = 0$ dBm, respectively.  Also,  we see that for $L\geq 100$, the SE with PSO for $\rho_d = -10$ dBm outperforms the SE with random phase shifts for $\rho_d = 0$ dBm. This illustrates the effectiveness of optimizing the RIS phase shifts.}

\vspace{-8pt}
\subsection{Impact of Shadow Fading Coefficient $\sigma_{sh}$ and $d_{\text{min}}$}
{We now in Fig. \ref{fig:10} plot SE versus AP transmit power considering three different values of shadow fading coefficient ($\sigma_{sh}$) i.e.,  $0$~dB, $4$~dB and $14$~dB for the direct links.  Typically,  a wireless system has a diverse set of users distributed randomly within its coverage area.  Within this setup,  there are users with both strong and weak links to APs.  Weak links can result from factors such as extended distances or significant obstructions causing large path loss and heavy shadowing.  We observe from Fig. \ref{fig:10} with the integration of RIS, we get a significant boost in the SE value.  For $\rho_d=10$~dBm,  the SE with RIS-assisted links increases by $12.5\%$ and $32\%$ for  $\sigma_{sh}=0$~dB and  $4$~dB  respectively,  as compared to the scenario when the RIS-assisted links are absent.  As the direct link further becomes weak i.e.,  for $\sigma_{sh}=14$~dB the impact of RIS on SE is more significant.  Therefore this study justifies the using the RIS when there is a weak direct link between the user and the AP.  We also observe that for higher transmit power, i.e., $\rho_d \geq 40$~dBm, the SE obtained with and without RIS are equal.  This is because at higher transmit powers, the received signal power for direct link becomes similar to that with RIS-aided~link.}

{We plot in Fig. \ref{fig:11} the SE versus AP transmit power for three different values of $d_\text{min}$. Remember the users are deployed outside a semi-circle of radius $d_\text{min}$~(in m) from the RIS. We observe from Fig. \ref{fig:11} that the SE decreases as the distance between RIS and users ($d_\text{min}$) increases, which is also expected as the channel between RIS and users weakens with the increase in distance.}

\begin{figure*}[htbp]
	\centering	
        \begin{subfigure}[b]{0.30\linewidth}              \hspace{-30pt}\includegraphics[width=1.2\linewidth,height=1\linewidth]{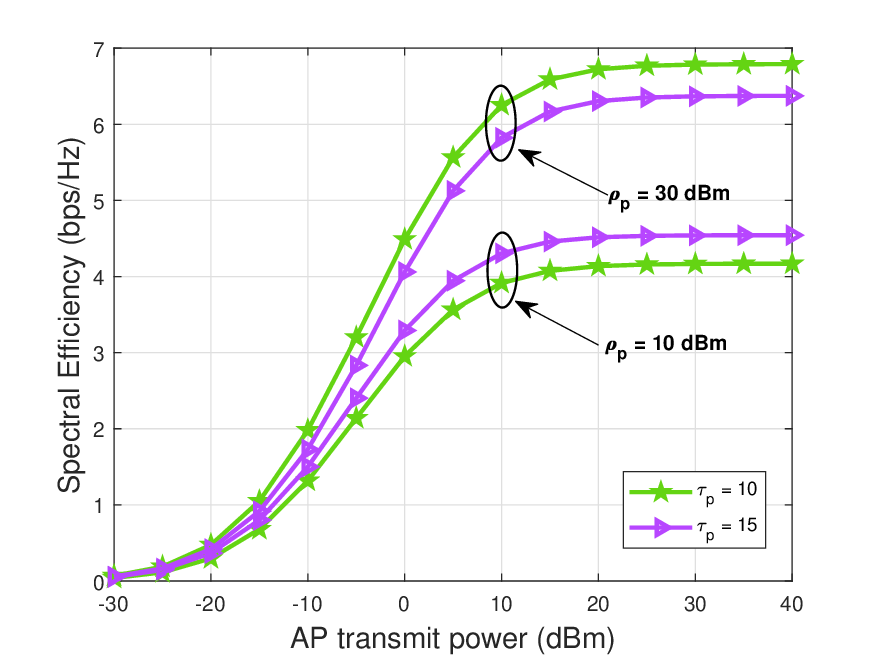}\vspace{-5pt}
                \caption{\small}
                \label{fig:13}
        \end{subfigure}\vspace{-3pt}
        \begin{subfigure}[b]{0.30\linewidth}	   \hspace{-15pt}\includegraphics[width=1.2\linewidth,height=1\linewidth]{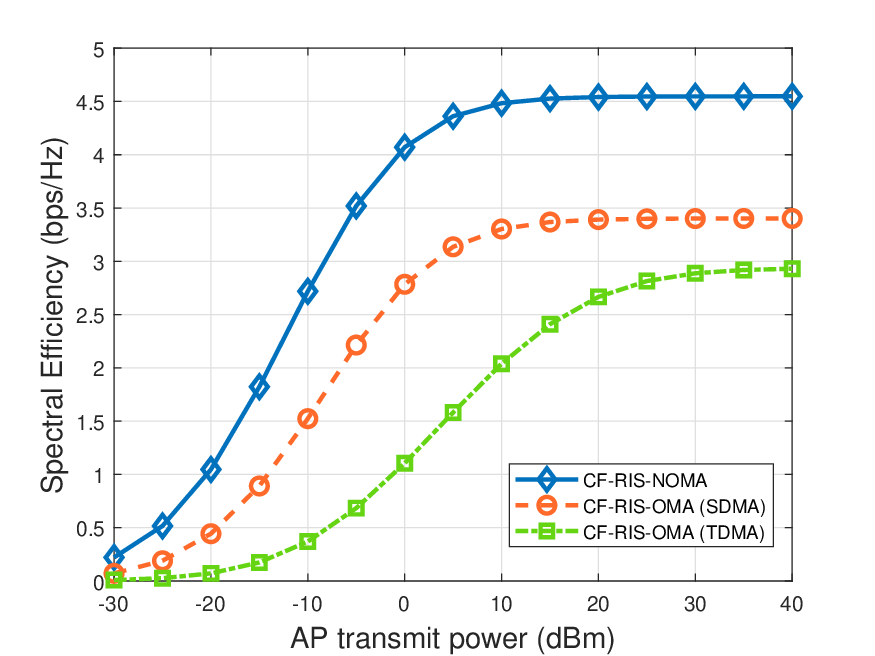}\vspace{-5pt}
              \caption{\small}
	        \label{fig:14}
	\end{subfigure}\vspace{-3pt} 
        \begin{subfigure}[b]{0.30\linewidth}	   \hspace{0pt}\includegraphics[width=1.2\linewidth,height=1\linewidth]{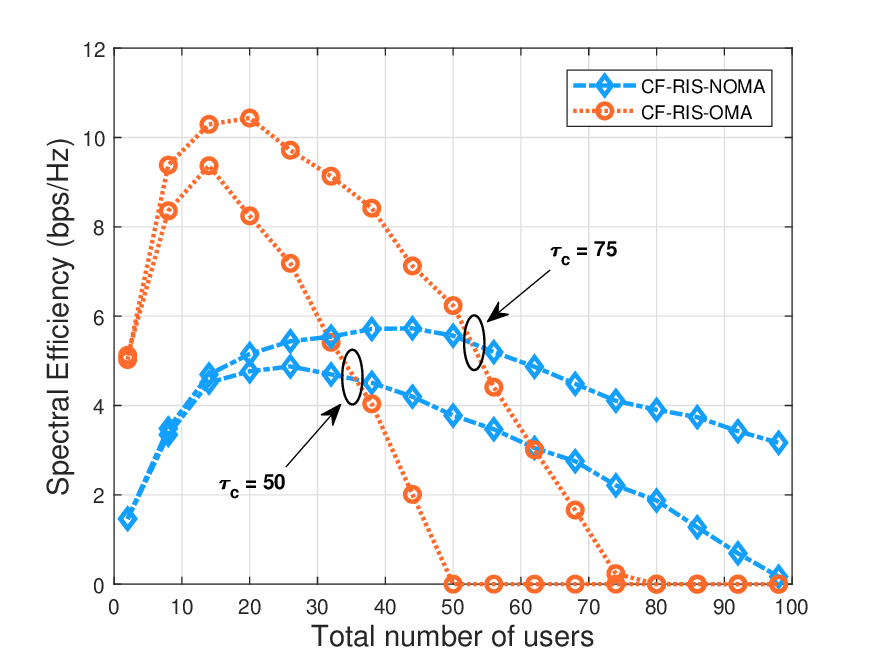}\vspace{-5pt}
              \caption{\small}
	       \label{fig:15}
	\end{subfigure} \vspace{-3pt} 
	\caption{\small {a) Spectral efficiency versus AP transmit power for different pilot duration, b) Spectral efficiency comparison between NOMA, SDMA and TDMA, c) Spectral efficiency versus total number of users for different coherence interval. \\[-30pt]}}
\end{figure*} 

\vspace{-10pt}
\subsection{{Convergence of the Proposed Algorithm}}
{To demonstrate the convergence of the proposed algorithm, we now plot in Fig.~\ref{fig:12} the SE versus the number of iterations required for the Algorithm~$3$ to converge,  by varying the AP transmit power.  For this study,  we set $M = 64$ APs, $K = 10$ clusters, $N = 2$ users per cluster and $L = 64$ RIS elements.  We observe from Fig.~\ref{fig:12} that Algorithm~$3$ requires a small number of iterations (i.e., around $25$ iterations) to converge.}

\vspace{-12pt}
\subsection{Impact of Pilot Length ($\tau_p$) and Pilot Power ($\rho_p$) on the Achievable SE}\vspace{-2pt}
{We now plot in Fig.~\ref{fig:13} the SE versus AP transmit power for two sets of pilot length (i.e., $\tau_p = 10$ and $15$) and pilot power (i.e., $\rho_p = 10$ and $30$ dBm).  For this study,  we set $M= 64$ APs, $L= 64$ RIS elements, $K= 10$ NOMA clusters and $N=2$ users in each cluster.  We observe from Fig.~\ref{fig:13} that for $\rho_p = 10$~dBm the SE improves as the pilot length is increased from~$10$ to $15$.  This is because the increase in pilot length results in better channel variance.  On the contrary,  for $\rho_p = 30$~dBm the SE degrades,  since the impact of increase in channel estimation overhead ($1- {\tau_p}/{\tau_c}$) dominates the improvement in channel variance.   Therefore,  the system designer should judiciously select the values of pilot power and pilot duration to achieve improved SE.} 

\vspace{-12pt}
\subsection{{Comparison of RIS Assisted Cell-Free NOMA with SDMA and TDMA}}\vspace{-2pt}
{We now in Fig.~\ref{fig:14} compare the SE obtained using the proposed RIS assisted cell-free NOMA with SDMA and time division multiple access (TDMA). For this study, we fix $M=64$ APs, $L=128$ RIS elements and a total $40$ number of users and vary the AP transmit power. In TDMA, each user is allocated a separate time slot. We observe that NOMA outperforms both SDMA and TDMA in terms of SE. In TDMA, the SE is limited by the fact that each user gets only a fraction of the total time to transmit \cite{TD3}. We also see for $\rho_d\geq10$ dBm, NOMA achieves a relative improvement of $28\%$ and $125\%$ in the SE as compared to SDMA and TDMA, respectively.}

\vspace{-12pt}
\subsection{SE versus Users for Different Coherence Interval} \vspace{-2pt}
{We now in Fig. \ref{fig:15} plot the SE versus number of users for two different coherence intervals i.e.,  $\tau_c=50$ and $\tau_c=75$.  We set $M= 64$ APs, $L= 128$ RIS elements, $N=2$ users in each NOMA cluster, and the maximum transmit power, $\rho_d= 20$~dBm. For a coherence interval of $\tau_c=50$,  we note that for OMA, the SE becomes zero when the total number of users reaches $50$. This is because, for OMA, the channel estimation overhead factor $\left(1- \frac{KN}{\tau_c}\right)$ becomes zero since each user is assigned with an orthogonal pilot.  Whereas for NOMA each cluster is assigned orthogonal pilot,  the channel estimation overhead factor $\left(1- \frac{K}{\tau_c}\right)$ becomes zero only when the number of NOMA clusters is equal to the  coherence interval.  Due to intra-cluster pilot contamination and imperfect SIC,  NOMA performs inferior to OMA when the number of users served is less than $36$ for $\tau_c=50$.  Similar observations can be made for $\tau_c=75$.}

\vspace{-10pt}
\section{Conclusion} \vspace{-2pt}
We derived a closed-form SE expression for a spatially correlated RIS-assisted cell-free mMIMO NOMA system with imperfect instantaneous CSI and imperfect SIC. We next formulated the joint power and phase optimization problem and proposed an alternate optimization algorithm to maximize the sum SE. We first employed a novel successive-QT algorithm to optimize the transmit power coefficients using block optimization along with QT, which decomposed the non-convex problem into convex sub-problems and then solved the RIS phase optimization problem using the PSO algorithm. We investigated the performance of the cell-free NOMA system and showed that the SE with the aid of RIS always performs better than the SE without RIS. Numerical investigations revealed that the presence of the RIS-assisted link is more useful, specifically at the lower transmit power region where the direct link from AP to user is weak. We also showed that the SE saturates at higher transmit powers and starts decreasing if the total number of users increases. We next compared our system with the OMA counterpart to show that the incorporation of NOMA results in better SE as compared to OMA. {Future work can consider modifications to the proposed cell-free massive MIMO system by incorporating an active RIS.} 

\vspace{-15pt}
\appendices
\section{} \label{A} 
The variance of the aggregated channel, $\delta_{mkn}$ in \eqref{c_mk} is computed first as  
\begin{align} \label{delta_mkn}
    \delta_{mkn} &= \mathbb{E}\left\{ \left|u_{mkn}\right|^2 \right\} = \mathbb{E}\left\{ \left| l_{mkn} + \mathbf{h}_{rkn}^H \mathbf\Theta_r \mathbf{g}_{mr}\right|^2 \right\}, \nonumber\\
     &\stackrel{(a)}{=} \beta_{mkn} + \mathbb{E}\left\{ \text{Tr}(\mathbf{\Theta}_{r} \mathbf{g}_{mr} \mathbf{g}_{mr}^{H} \mathbf{\Theta}_{r}^{H} \mathbf{h}_{rkn} \mathbf{h}_{rkn}^H ) \right\},\nonumber\\
     &= \beta_{mkn} + \text{Tr}\left( \mathbf{\Theta}_{r} \mathbf{R}_{mr} \mathbf{\Theta}_{r}^{H} \mathbf{R}_{rkn}\right).
\end{align}
Equality $(a)$ is obtained using the property $\text{Tr}(\text{XY}) = \text{Tr}(\text{YX})$ for some given size-matched matrices $X$ and $Y$, and the fact that the channels $\mathbf{h}_{rkn}$ and $\mathbf{g}_{mr}$ are statistically independent. Using \eqref{DS_BU_ICI}, the desired signal term $\text{DS}_{kn}$ in~\eqref{Gamma} can be simplified~as 
\begin{align} \label{DS_appendix}
\text{DS}_{kn}\; &{=} \left| \mathbb{E}\left\{ \sum\limits_{m = 1}^M \sqrt{\eta_{mkn} \rho_d} {u}_{mkn} {\hat{z}}_{m k}^{*}  \right\}\right|^2, \nonumber\\
&\stackrel{(a)} {=} \left|\sum\limits_{m = 1}^M \mathbb{E}\left\{ \sqrt{\eta_{mkn} \rho_d} u_{mkn} c_{mk}\left( \sqrt{\tau_p \rho_p} \sum_{p=1}^N u_{mkp}  \right) \right\}\right|^2\!\!\!, \nonumber\\
&\stackrel{(b)} {=}\bigg( \sum_{m=1}^M  {\sqrt{\eta_{mkn} \rho_d} \gamma_{mk} \frac{\delta_{mkn}}{\sum_{n'=1}^N \delta_{mkn'}}  }\bigg)^2\!,
\end{align}
where equality $(a)$ is obtained by first substituting the precoder, $\hat{z}_{mk}=\sum_{i=1}^N {u}_{mki}$ and equality $(b)$ by using the fact that the channel estimation error and the channel estimate are uncorrelated to each other. 
We next evaluate the beamforming uncertainty gain $\text{BU}_{kn}$ in \eqref{DS_BU_ICI} as follows
\begin{align*}
\text{BU}_{kn} \!
&= {\rho_d}\mathbb{E}\!\left| \sum\limits_{m=1}^M \!\!\!\sqrt{\eta_{mkn}} {u}_{mkn} {\hat{z}}_{mk}^* - \mathbb{E} \!\left\{\!\sum\limits_{m=1}^{M} \!\!\!\sqrt{\eta_{mkn}} {u}_{mkn} {\hat{z}}_{mk}^*\! \right\}\!\right|^2\!\!\!,\nonumber \\ 
&= \rho_d \sum\limits_{m=1}^M \eta_{mkn}\left( \underbrace{\mathbb{E} \left\{ \left| {u}_{mkn} {\hat{z}}_{mk}^* \right|^2 \right\}}_{A}-  \underbrace{\left| \mathbb{E}\left\{ {u}_{mkn} {\hat{z}}_{mk}^* \right\}\right|^2}_B \right)\!\!,
\end{align*}
where $A$ can be evaluated as $A \stackrel{(a)}{=}\tau_p \rho_p c_{mk}^2 \left(\mathbb{E}\left\{|u_{mkn}|^4\right\} + \frac{1}{\tau_p \rho_p}\delta_{mkn} + C \right)$, here equality $(a)$ is obtained by decomposing the expression into fourth order moment of the aggregated channel, variance of the aggregated channel and $C = \mathbb{E}\left\{\left|\sum_{p \ne n}^N u_{mkp}^*\right|^2\right\}$. Furthermore, the term $C$ can be evaluated as $ C =  \left(\frac{1}{\tau_p \rho_p c_{mk}^2}\delta_{mkn}\gamma_{mk} - \frac{1}{\tau_p \rho_p}\delta_{mkn} - \delta_{mkn}^2 \right)$. The term $B$ can be calculated from the same identity used in calculating the desired signal term in \eqref{DS_appendix}. After adding all these terms, the power of the beamforming uncertainty can be simplified~as 
\begin{align} \label{BU_appendix}
\text{BU}_{kn} = \rho_d \sum\limits_{m=1}^M {{\eta_{mkn} }} \delta_{mkn} \gamma_{mk}.
\end{align}
We next compute the inherent intra-cluster interference $\text{IaCI}^{I}_{kn'}$ in~\eqref{DS_BU_ICI} as follows  
\begin{align*}
\text{IaCI}^{I}_{kn'} = \sum_{n'=1}^{n-1} \underbrace{\mathbb{E} \left\{\left| \sum\limits_{m = 1}^M \sqrt{\eta_{m k n'} \rho_d} {u}_{mkn} {\hat{z}}_{mk}^{*}  \right|^2\right\}}_{{T}}.
\end{align*} 
Now the expectation term inside the summation can be decomposed into two terms as  
\begin{align*}
T\;\stackrel{(a)}{=}& \underbrace{\left|\sum\limits_{m=1}^M {\sqrt{ \eta_{mkn'}\rho_d}} c_{mk} \mathbb{E} \left\{  {u}_{mkn}  {\hat{z}}_{mk}^*  \right\}  \right|^2}_{{T_1}} \\ \nonumber
&{+}\underbrace{\sum\limits_{m=1}^M {\eta_{mkn'}\rho_d} c_{mk} \mathbb{E}\left\{ \left| {u}_{mkn}{\hat{z}}_{mk}^* - \mathbb{E}\left\{ {u}_{mkn} {\hat{z}}_{mk}^*  \right\} \right|^2\right\}}_{{T_2}}. 
\end{align*} 
We can calculate the term $T_1$ as 
\begin{align*}
    T_1\;&\stackrel{(b)}{=} \left|\sum\limits_{m=1}^M \sqrt{\eta_{mkn'}\rho_d} \mathbb{E} \left\{ \left| {\hat{u}}_{mkn} {\hat{z}}_{mk}^* + {e}_{mkn} {\hat{z}}_{mk}^* \right| \right\} {}\right|^2, \nonumber \\
    &\stackrel{(c)}{=} \left(\sum\limits_{m=1}^M {\sqrt{\eta_{mkn'}\rho_d}} \gamma_{mk}^2 \frac{\delta_{mkn}}{\sum_{p=1}^N \delta_{mkp}} \right)^2,
\end{align*} 
where equality $(b)$ follows the relation between channel estimation and the channel estimation error, and $(c)$ follows the derivation similar to \eqref{DS_appendix}. Similarly, we can calculate $T_2$ as 
\begin{align*}
T_2\;&\stackrel{(d)}{=} \sum\limits_{m=1}^M  {{\eta_{mkn'}\rho_d}} \left(\mathbb{E} \left\{ \left| {\hat{u}}_{mkn} \right|^4 \right\} + \mathbb{E} \left\{\left| {\hat{u}}_{mkn}^* {e}_{mkn}  \right|^2 \right\} \right), \nonumber \\
&\stackrel{(e)}{=}\sum\limits_{m=1}^M {{\eta_{mkn'}\rho_d}}  \gamma_{mk} \delta_{mkn},
\end{align*}
where equality $(d)$ follows the same identity used in calculating the beamforming uncertainty in \eqref{BU_appendix}. Equality $(e)$ uses the fact that the channel estimation error and the channel estimate are uncorrelated to each
other. Using the final expressions of $T_1$ and $T_2$, we can write the expression for inherent intra-cluster interference as 
\begin{align}\label{IaCi_appendix}
\text{IaCI}^{I}_{kn'} \!\!=\!\! \sum_{n'=1}^{n-1}\!\!\rho_d \bigg(\!\left|\sum\limits_{m=1}^M \frac{{\sqrt{\eta_{mkn'}}} \gamma_{mk}\delta_{mkn}}{\sum_{p=1}^N \delta_{mkp}} \right|^2 \!\!\!\!+\!\! \sum\limits_{m=1}^M \!\!{{\eta_{mkn'}}}  \gamma_{mk} \delta_{mkn}\! \bigg).
\end{align} 
We now evaluate the residual intra-cluster interference $\text{IaCI}^{R}_{kn'}$ in~\eqref{DS_BU_ICI} as follows 
\begin{align*}
& \text{IaCI}^{R}_{kn'}\nonumber\\
 & \!=\!  \sqrt{\rho_d}\!\!\! \sum_{n'=n+1}^{N}\!\!\!\! \bigg(\!\sum\limits_{m = 1}^M \!\!\sqrt{\eta_{mkn'}} u_{mkn} \hat{z}_{mk}^* \!-\!\! 
      \mathbb{E}\!\left\{ \sum_{m=1}^M \!\!\sqrt{\eta_{mkn'}} u_{mkn} \hat{z}_{mk}^* \!\!\right\}\!\!\bigg).\!\!
\end{align*} 
Substituting the precoder $\hat{z}_{mk}= \sum_{p=1}^N u_{mkp}$ we get
\begin{align}\label{IaCr_appendix}
&\text{IaCI}^{R}_{kn'}  
\stackrel{(a)}{=} \sum_{n'=n+1}^{N}\sum\limits_{m=1}^M {{ \eta_{mkn'}\rho_d}} \delta_{mkn} \gamma_{mk},
\end{align}
where equality $(a)$ follows algebraic manipulations and assumes that the actual and estimated channel behavior is independent.
Finally, we calculate the inter-cluster interference $\text{ICI}_{k'n'}$ in \eqref{DS_BU_ICI} as follows 
\begin{align*}
&\text{ICI}_{k'n'} =  \sum_{k'\neq k}^K\sum_{n'=1}^N\sum\limits_{m=1}^M {\eta_{mk'n'}\rho_d} {\mathbb{E}\left\{ \left|{u}_{mkn}{\hat{z}}_{mk'}^* \right|^2\right\}}, \\
&\quad=\tau_p \rho_p c_{mk}^2\!\left(\! \mathbb{E} \left\{ \left| {u}_{mkn} \sum_{p=1}^N \!u_{mkp}^*\right|^2 \!\right\} \!+\! \mathbb{E} \!\left\{\! \left| {u}_{mkn}\mathbf{\tilde{w}}_{pm} \right|^2 \!\right\}\!\!\right)\!. 
\end{align*}
Simplifying the above expression, we can write
\begin{align}\label{ICI_appendix}
& \text{ICI}_{k'n'} = \sum_{k'\neq k}^K\sum_{n'=1}^N\sum_{m=1}^M {\eta_{mk'n'}\rho_d} \delta_{mkn} \gamma_{mk'} .
\end{align}
The proof follows by substituting \eqref{DS_appendix}-\eqref{ICI_appendix} into \eqref{Gamma} and by using some algebraic manipulations.

\bibliographystyle{IEEEtran} 
\bibliography{cite}

\end{document}